\documentclass[journal,transmag]{IEEEtran}
\usepackage{color}

\usepackage{hyperref}

\ifCLASSINFOpdf
   \usepackage[pdftex]{graphicx}
\else
\fi
\usepackage{amsmath}
\begin{document}
%
\title{Efficient calculation of self magnetic field,  self-force, and self-inductance for electromagnetic coils. II. Rectangular cross-section}


\author{\IEEEauthorblockN{Matt Landreman\IEEEauthorrefmark{1},
Siena Hurwitz\IEEEauthorrefmark{1}, and
Thomas M Antonsen, Jr.\IEEEauthorrefmark{1}}
\IEEEauthorblockA{\IEEEauthorrefmark{1}Institute for Research in Electronics and Applied Physics, University of Maryland, College Park, MD 21043 USA}
\thanks{Corresponding author: M. Landreman (email: mattland@umd.edu).}}

%



\IEEEtitleabstractindextext{%
\begin{abstract}
For designing high-field electromagnets, the Lorentz force on coils must be computed to ensure a support structure is feasible, and the inductance should be computed to evaluate the stored energy.
Also, the magnetic field and its variation inside the conductor is of interest for computing stress and strain, and due to superconducting quench limits.
For these force, inductance, energy, and internal field calculations, the coils cannot be naively approximated as infinitesimally thin filaments due to divergences when the source and evaluation points coincide, so more computationally demanding calculations are usually required, resolving the finite cross-section of the conductors.
Here, we present a new alternative method that enables  the internal magnetic field vector, self-force, and self-inductance to be computed rapidly and accurately within a 1D filament model.
The method is applicable to coils for which the curve center-line can have general noncircular shape, as long as the conductor width is small compared to the radius of curvature.
This paper extends a previous calculation for circular-cross-section conductors [Hurwitz et al, arXiv:2310.09313 (2023)] to consider the case of rectangular cross-section.
The reduced model is derived by rigorous analysis of the singularity, regularizing the filament integrals such that they match the true high-dimensional integrals at high coil aspect ratio.
The new filament model exactly recovers analytic results for a circular coil, and is shown to accurately reproduce full finite-cross-section calculations for a non-planar coil of a stellarator magnetic fusion device.
Due to the efficiency of the model here, it is well suited for use inside design optimization.
\end{abstract}

}

\maketitle

\IEEEdisplaynontitleabstractindextext

%
\IEEEpeerreviewmaketitle

\section{Introduction}
%
%
%
%
\IEEEPARstart{I}{n} 
a configuration of electromagnets, important quantities include the Lorentz forces, inductances, and magnetic fields inside the conductors.
The forces determine the necessary support structure, and the inductances determine the stored magnetic energy and dynamics.
The magnetic field in the conductor determines local forces and their shear, and in superconductors the field also determines the critical current.
For general steady currents, the field is determined via a three-dimensional (3D) Biot-Savart integral,
\begin{equation}
    \mathbf{B}(\mathbf{r}) = \frac{\mu_0}{4\pi}
    \int d^3\tilde{r} \frac{\mathbf{J}(\tilde{\mathbf{r}}) \times (\mathbf{r} - \tilde{\mathbf{r}})}{\left| \mathbf{r}-\tilde{\mathbf{r}}\right|^3},
    \label{eq:BiotSavart_3D}
\end{equation}
where $\mathbf{J}$ is the current density.
Similarly, the force per unit length $d\mathbf{F}/d\ell$ is determined by a 5D integral, given by integrating (\ref{eq:BiotSavart_3D}) over the conductor's cross-sectional area, 
and the self-inductance $L$ is given by a 6D integral:
\begin{equation}
    L = \frac{\mu_0}{4\pi I^2}
    \int d^3 r \int d^3\tilde{r} \frac{\mathbf{J}(\mathbf{r}) \cdot \mathbf{J}(\tilde{\mathbf{r}}) }{\left| \mathbf{r}-\tilde{\mathbf{r}}\right|},
    \label{eq:inductance_6D}
\end{equation}
where $I$ is the current, and the stored energy is $W = (1/2)LI^2$.
If one considers two thin coils $A$ and $B$ that are separated by more than the conductor width, the effect of one on the other can be computed by approximating them as infinitesimally thin filaments.
The field and force at a point $\mathbf{r}_A$ reduce to 1D integrals over  coil $B$,
\begin{equation}
\label{eq:Biot_savart_1D}
    \mathbf{B}(\mathbf{r}_A) = \frac{\mu_0 I_B}{4\pi}
    \int \frac{d{\mathbf{r}_B} \times (\mathbf{r}_A - {\mathbf{r}_B})}{\left| \mathbf{r}_A-{\mathbf{r}_B}\right|^3},
\end{equation}
\begin{equation}
    \left(\frac{d\mathbf{F}}{d\ell}\right)_A 
    = \frac{\mu_0 I_A I_B}{4\pi}
    \mathbf{t}_A \times
    \int \frac{d{\mathbf{r}_B} \times (\mathbf{r}_A - {\mathbf{r}}_B)}{\left| \mathbf{r}_A-{\mathbf{r}_B}\right|^3},
\end{equation}
where $\mathbf{t}$ is the unit tangent vector.
Similarly, the mutual inductance $M$ can be computed with a 2D integral over both coils:
\begin{equation}
\label{eq:mutual_inductance}
    M = \frac{\mu_0}{4\pi}
    \iint \frac{d\mathbf{r}_A \cdot d{\mathbf{r}}_B}{\left| \mathbf{r}_A-{\mathbf{r}_B}\right|}.
\end{equation}
This reduction in dimensionality is extremely advantageous for simplicity and speed of computation.
The self-field, self-force, and self-inductance of a coil cannot be computed by setting $B \to A$ in (\ref{eq:Biot_savart_1D})-(\ref{eq:mutual_inductance}) due to the singularity in the denominators when the source and evaluation points coincide.
However, in this paper we will derive modified singularity-free 1D integrals for efficient calculation of the self-field and self-force, and a 2D integral for calculation of the self-inductance.
The new formulas here are derived by rigorous analysis of the region where source and evaluation points nearly coincide, resolving the singularities in the filamentary integrals appropriately to have the correct asymptotic behavior.
For one example documented in this paper, calculating the self-force per unit length using the reduced model derived here is $\sim$ 18,000 times faster than using the original high-dimensional integral.

\begin{figure}[!t]
\centering
\includegraphics[height=3.7in]{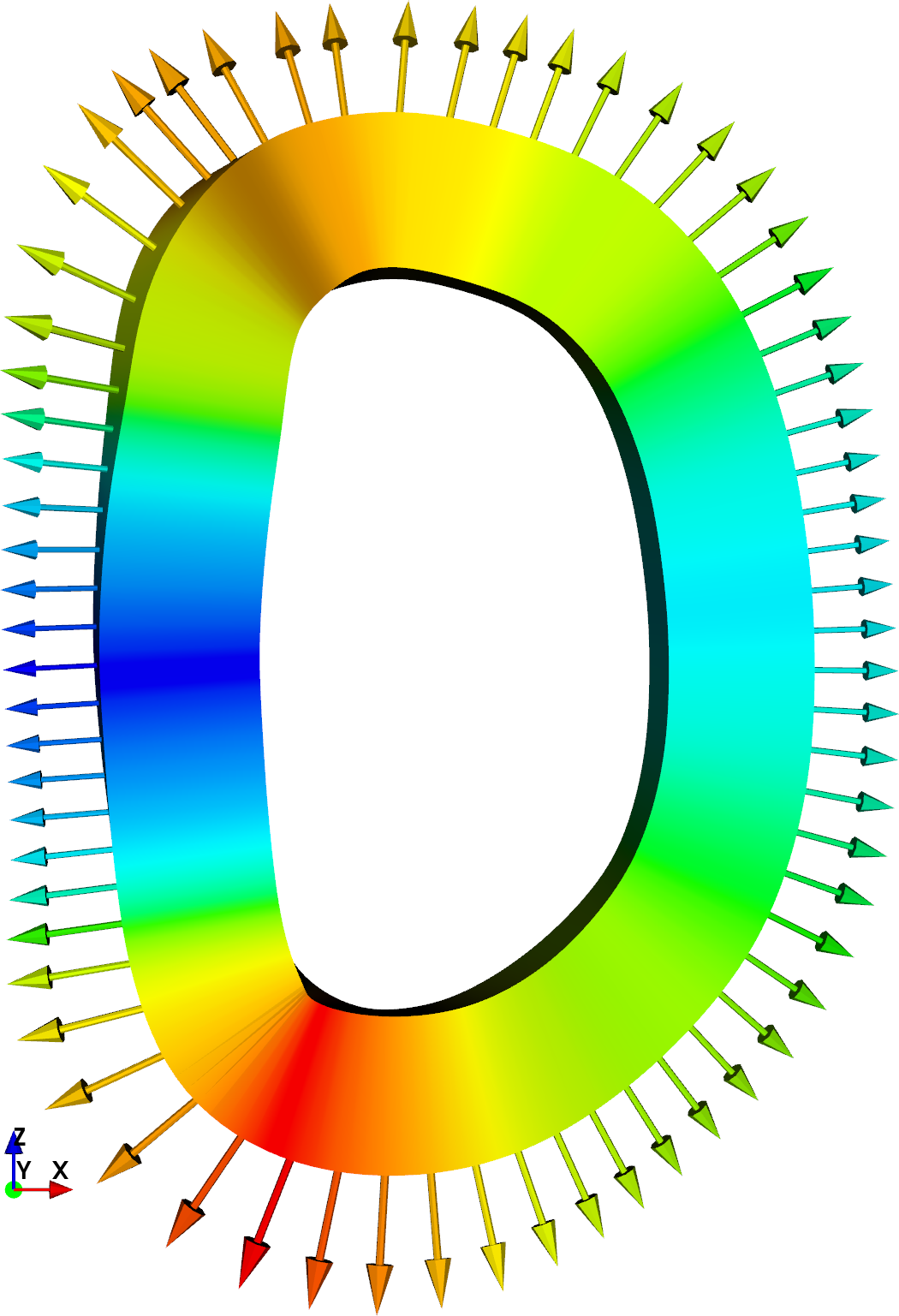}
\includegraphics[height=3.7in]{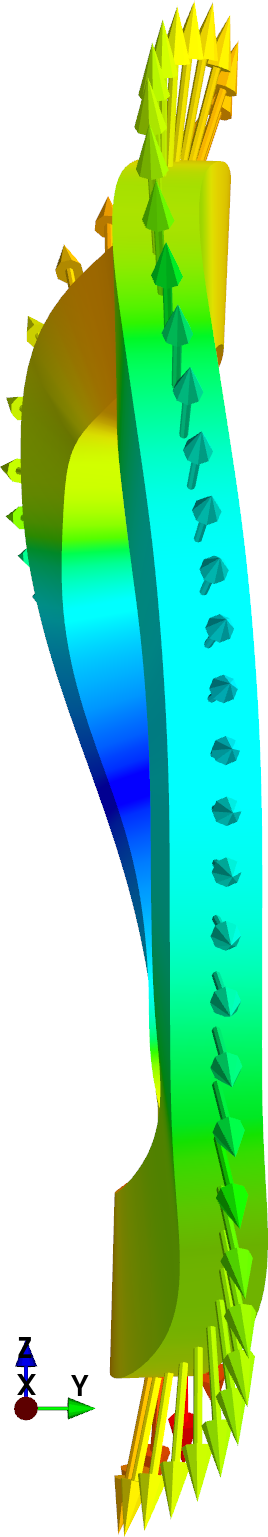}
\caption{Coil from the HSX stellarator, shown from two angles. 
The cross-section dimensions $a$ and $b$ shown are those of the experiment.
Vectors and color indicate the self-force, averaged over the cross-section.}
\label{fig:3D}
\end{figure}

One motivating application for our work is the design of electromagnets for magnetic confinement fusion.
These magnets have a large bore to enclose the plasma that they confine.
Confinement of the plasma improves with field strength \cite{whyte2019small}, so high fields are employed, and supporting the large Lorentz forces is a serious concern \cite{panin2016approaches}.
The reduction of the critical current for superconductivity due to the large field is also important.
Figure \ref{fig:3D} shows an illustrative example of a fusion-relevant magnet, one of the non-planar coils for the Helically Symmetric eXperiment (HSX) at the University of Wisconsin \cite{HSX}.
The shapes of fusion magnets, particularly for the stellarator concept, may be designed by optimization.
It is therefore valuable to compute the Lorentz force and other quantities rapidly inside the optimization loop.
The results in this paper are also applicable to the many other contexts in which shapes of electromagnets are optimized.
These applications include magnetic resonance imaging \cite{hidalgo2010theory, chen2017electromagnetic} and  accelerator beam optics \cite{russenschuck2011field}.

\begin{figure}[!t]
\centering
\includegraphics[width=\columnwidth]{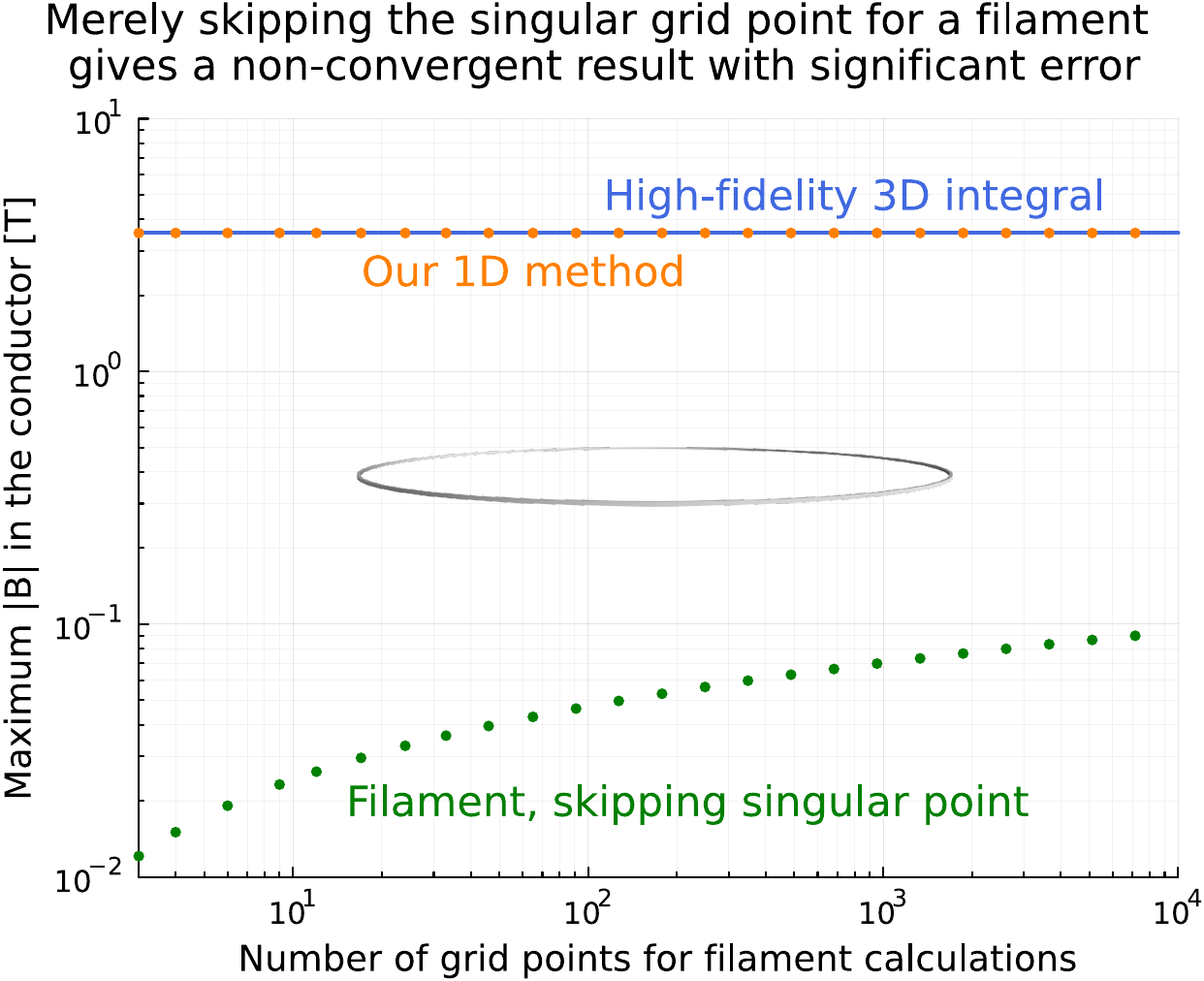}
\caption{The field strength on the inner edge of a circular coil (inset) with major radius 1 m, square cross-section of size 1 cm $\times$ 1 cm, and 100 kA current, computed three ways.
Even with few grid points, our reduced model agrees with a high-fidelity calculation.
A simplistic 1D filament model in which the singular point is ignored is highly inaccurate, not converging to any limit.
}
\label{fig:non_convergence}
\end{figure}

One might hope that the singularities in the filamentary integrals (\ref{eq:Biot_savart_1D})-(\ref{eq:mutual_inductance}) with $B \to A$ could be resolved in numerical calculations by merely neglecting the singular quadrature point \cite{gatto2023time}.
However, this approach leads to substantial errors in the self-field, self-force, and self-inductance.
The results of such calculations diverge logarithmically as the number of grid points increases, rather than converging to any limit.
Consequently, analytic results for geometries such as a circular coil cannot be reproduced.
These problems with simply dropping the singular grid point are demonstrated in figure 3 of \cite{Hurwitz} for the force, and in figure \ref{fig:non_convergence} here for the field.

Proper methods for calculating the self-field, self-force, and/or self-inductance using reduced models have been previously proposed by several authors.
For coils with rectangular cross-section composed of straight sections and circular arcs, formulas for the internal field have been derived analytically \cite{sackett1975calculation, urankar1982vector, Lion}.
However reductions of these formulas to a 1D integral for the force or 2D integral for the inductance and energy do not appear to have been reported.
Also, many coil shapes of interest are not composed of straight sections and circular arcs, such as the one in figure \ref{fig:3D}.
In \cite{RobinVolpe}, the self-force was evaluated for a sheet current, which may be a reasonable approximation when there are a large number of adjacent filaments.
Garren \& Chen \cite{GarrenChen} evaluated the self-force on current filaments of general shape, motivated by the effect of self-force on driving solar flare eruptions.
In that work, a method to compute the force using only a 1D integral was proposed in which a specific interval of the integration domain was removed, avoiding the singularity.
Dengler \cite{Dengler} proposed a similar method of removing a specific interval from 2D integrals for computing self-inductance.
These last three works did not specifically consider the magnetic field inside the conducting filament.

In the present paper, we derive computationally efficient reduced integrals for the self-field, self-force and self-inductance with several advantages compared to these previous approaches.
All three of these quantities are computed in a unified framework.
The coil geometry is allowed to be quite general, not necessarily composed of straight sections and circular arcs.
We assume only that the cross-section is rectangular, with dimensions that do not vary along the coil, and that the radius of curvature of the coil is large compared to the cross-section dimensions (disallowing sharp corners).
No assumption is made about the rectangle's orientation or its rotation along the coil.
Unlike the approaches in \cite{GarrenChen, Dengler}, our reduced integrals cover the full periodic domain along the coil.
This means that the simplest possible quadrature grid, points uniformly spaced in the curve parameter, gives spectrally accurate integrals.
Also, data such as the position vector and curvature from each quadrature point can naturally be re-used to compute the self-field and self-force at the other quadrature points.
In contrast, the approaches in \cite{GarrenChen, Dengler} require custom quadrature grids for evaluating the self-field or self-force at different points along the coil.

The reduced integrals derived here resemble the integrals one would get by assuming an infinitesmally thin coil (setting $B \to A$ in (\ref{eq:Biot_savart_1D})-(\ref{eq:mutual_inductance})), except that an extra term is present in the denominator, so there is no division by zero.
The extra term is negligible in most of the integration domain, but it becomes significant when the distance between source and evaluation points is small, comparable to the coil thickness.
Physically, this term reflects the fact that when the source and evaluation points are this close, the finite thickness of the conductor cannot be neglected in the true high-dimensional integrals.
In the reduced integrals, a factor of order unity is present multiplying the regularization term, and it is calculated rigorously to match the behavior of the true high-dimensional integrals.

This paper builds on our previous work in \cite{Hurwitz} for thin coils with circular cross-section.
In the present work, similar techniques are applied to coils where the cross-section is rectangular.
The case of rectangular cross-section is more complicated because the orientation of the cross-section must be specified, affecting some aspects of the calculations.
Many details of the calculation here differ from and are more complicated than the circular-cross-section case, although the final results have similarities in form.

One important assumption in our method is that the current density is uniform across the conductor cross-section.
At a coarse-grained level, this model of the current distribution is accurate for a coil composed of multiple turns in both the transverse directions, as each turn carries the same current.
This model may not be highly accurate for high temperature superconducting (HTS) tapes, in which case the current may distribute itself non-uniformly in the transverse dimension along the tape \cite{rostila2007self, gomory2006self}.
The uniform-current-density model also does not account for skin currents associated with diamagnetism that may be relevant for some superconductors.
Nonetheless, the model here may still be useful for superconducting coils for initial estimations or use inside design optimization.
A uniform current density approximation has been used for superconducting coils by other authors coils, e.g. section 2.2 of \cite{riva2023development}.

In the following section, the problem geometry is defined and coordinates are established.
Then in section \ref{sec:results}, the main results are collected together for convenience.
The methodology for deriving these results is explained in more depth in section \ref{sec:methodology}.
Sections \ref{sec:inductance}-\ref{sec:force} give specific details of the self-inductance, self-field, and self-force calculations respectively.
The reduced integrals have fine-scale structure in the integrand, and in section \ref{sec:singularity_subtraction} we show how to remove this structure analytically so the integrals can be evaluated accurately on coarse grids.
Finally the results are discussed in section \ref{sec:conclusion}, where we conclude.

\section{Problem geometry}

\begin{figure}[!t]
\centering
\includegraphics[width=\columnwidth]{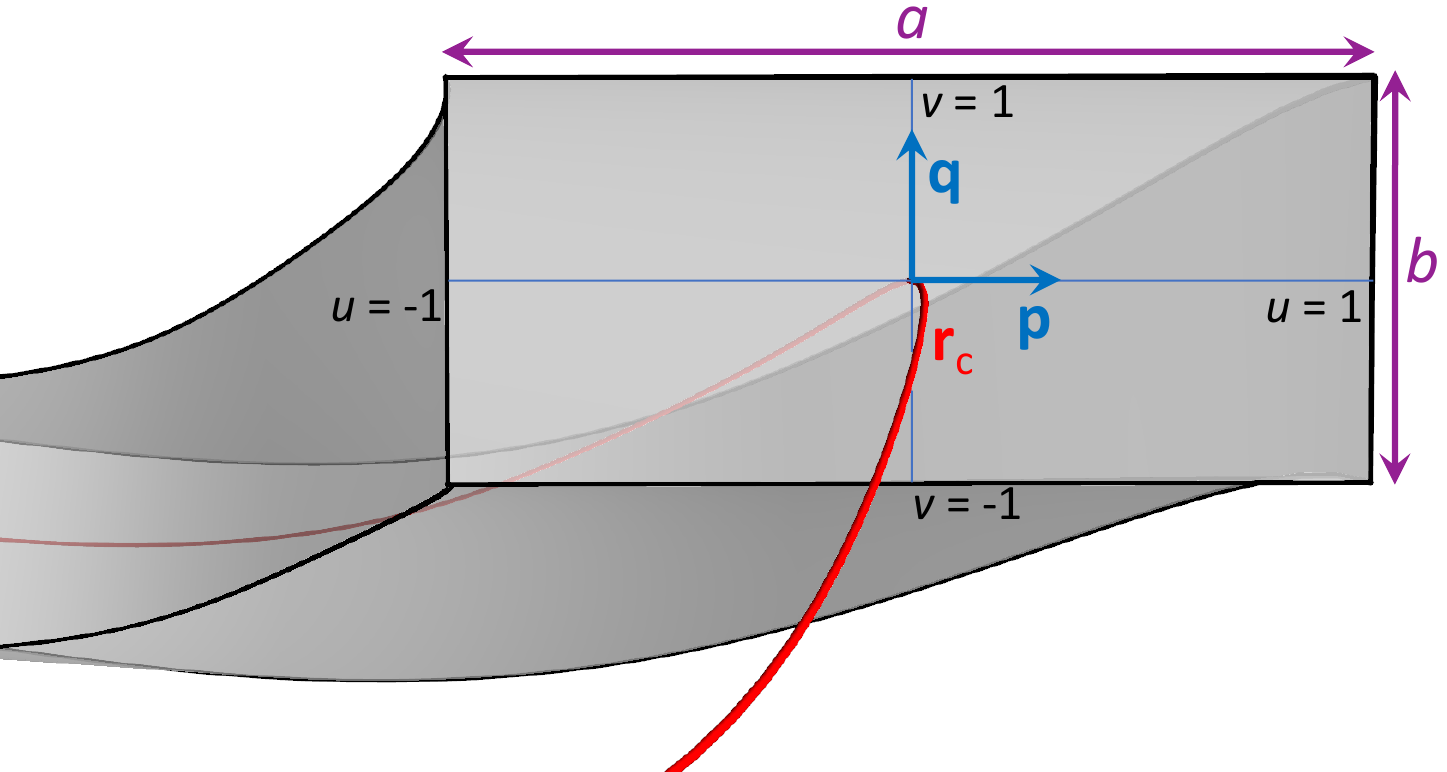}
\caption{Geometry definitions.
The coil center-line $\mathbf{r}_c$ is the curve shown in red.
A section of the coil conductor surrounding the center-line is shown in gray.
}
\label{fig:geometry}
\end{figure}

The conductor is considered to have a rectangular cross-section with side dimensions $a$ and $b$ at all points along its length.
We define coordinates to parameterize the conductor volume.
A periodic angle-like coordinate $\phi\in[0,2\pi)$ will be used parallel to the conductor, while coordinates $u, v \in [-1, 1]$ will be used in the transverse directions.
The position vector $\mathbf{r}$ is expressed as
\begin{equation}
\mathbf{r}(\phi,u,v) = \mathbf{r}_c(\phi)
+\frac{ua}{2}\mathbf{p}(\phi)
+\frac{vb}{2}\mathbf{q}(\phi),
\label{eq:position}
\end{equation}
where $\mathbf{r}_c(\phi)$ is the position vector of the coil center-line, and $\mathbf{p},\mathbf{q}$ are unit vectors in the perpendicular plane aligned with the conductor cross-section.
These definitions are illustrated in figure \ref{fig:geometry}.
The goal of our approach will be to express the inductance, self-force, and field strength in terms of integrals over $\phi$ involving $\mathbf{r}_c$.

The unit tangent of the center-line is $\mathbf{t} = |\mathbf{r}_c'|^{-1} \mathbf{r}_c'$, where primes denote $d/d\phi$. 
The set $\{\mathbf{t}, \mathbf{p}, \mathbf{q} \}$ forms an orthonormal curve frame for $\mathbf{r}_c$.
We consider this frame to be right-handed: $\mathbf{t}\cdot\mathbf{p}\times\mathbf{q} = 1$.
Differentiating $\mathbf{p}\cdot\mathbf{p}=1$, $\mathbf{p}\cdot\mathbf{q}=0$, etc., one finds \cite{Bishop}
\begin{equation}
    \frac{d}{d\phi}
    \begin{pmatrix}
    \mathbf{t} \\
    \mathbf{p} \\
    \mathbf{q}
    \end{pmatrix}
    = 
    \left| \frac{d\mathbf{r}_c}{d\phi}\right|
    \begin{pmatrix}
        0 && \kappa_1 && \kappa_2 \\
        -\kappa_1 && 0 && \kappa_3 \\
        -\kappa_2 && -\kappa_3 && 0
    \end{pmatrix}
   \begin{pmatrix}
    \mathbf{t} \\
    \mathbf{p} \\
    \mathbf{q}
    \end{pmatrix},
\label{eq:Frenet}
\end{equation}
for some functions $\{ \kappa_1(\phi), \kappa_2(\phi), \kappa_3(\phi) \}$. The standard curvature $\kappa(\phi)$ of the center-line is the magnitude of the derivative of $\mathbf{t}$ with respect to arclength, hence $\kappa=\sqrt{\kappa_1^2 + \kappa_2^2}$.
Noting the top row of (\ref{eq:Frenet}) is also equal to $\kappa\mathbf{n}|\mathbf{r}_c'|$, where $\mathbf{n}$ is the unit normal, then $\kappa_1\mathbf{p}+\kappa_2\mathbf{q}=\kappa\mathbf{n}$.
Crossing with $\mathbf{t}$ gives
\begin{equation}
    \kappa_1\mathbf{q}-\kappa_2\mathbf{p}=\kappa\mathbf{b},
    \label{eq:kappa_b}
\end{equation}
where $\mathbf{b}=\mathbf{t}\times\mathbf{n}$ is the binormal.
To compute $\kappa_1$ and $\kappa_2$, one can use $\kappa_{1}=\kappa\mathbf{n}\cdot\mathbf{p}=\kappa\mathbf{b}\cdot\mathbf{q}$, and $\kappa_{2}=\kappa\mathbf{n}\cdot\mathbf{q}=-\kappa\mathbf{b}\cdot\mathbf{p}$.
Note that since the $\{\mathbf{t},\mathbf{p},\mathbf{q}\}$ curve frame is defined to align with the rectangular conductor cross-section, the frame is smooth at any points or segments where $\kappa=0$, in contrast to the Frenet frame.


From (\ref{eq:position})-(\ref{eq:Frenet}), the Jacobian of the $(\phi,u,v)$ coordinates is
\begin{equation}
    \sqrt{g} = \frac{ab \left| \mathbf{r}_c'\right|}{4}  \left( 1-\frac{\kappa_1 au}{2} -\frac{\kappa_2 bv}{2} \right).
\label{eq:Jacobian_exact}
\end{equation}

We consider the current density to be
\begin{eqnarray}
    \mathbf{J}=\frac{I}{ab}\mathbf{t},
    \label{eq:J}
\end{eqnarray}
where $I$ is the total current, i.e. uniform in magnitude throughout the conductor.
Using the formula for the divergence in curvilinear coordinate systems, it can be verified that (\ref{eq:J}) implies $\nabla\cdot\mathbf{J}=0$.


\section{Main results}
\label{sec:results}

\begin{figure}[!t]
\centering
\includegraphics[width=\columnwidth]{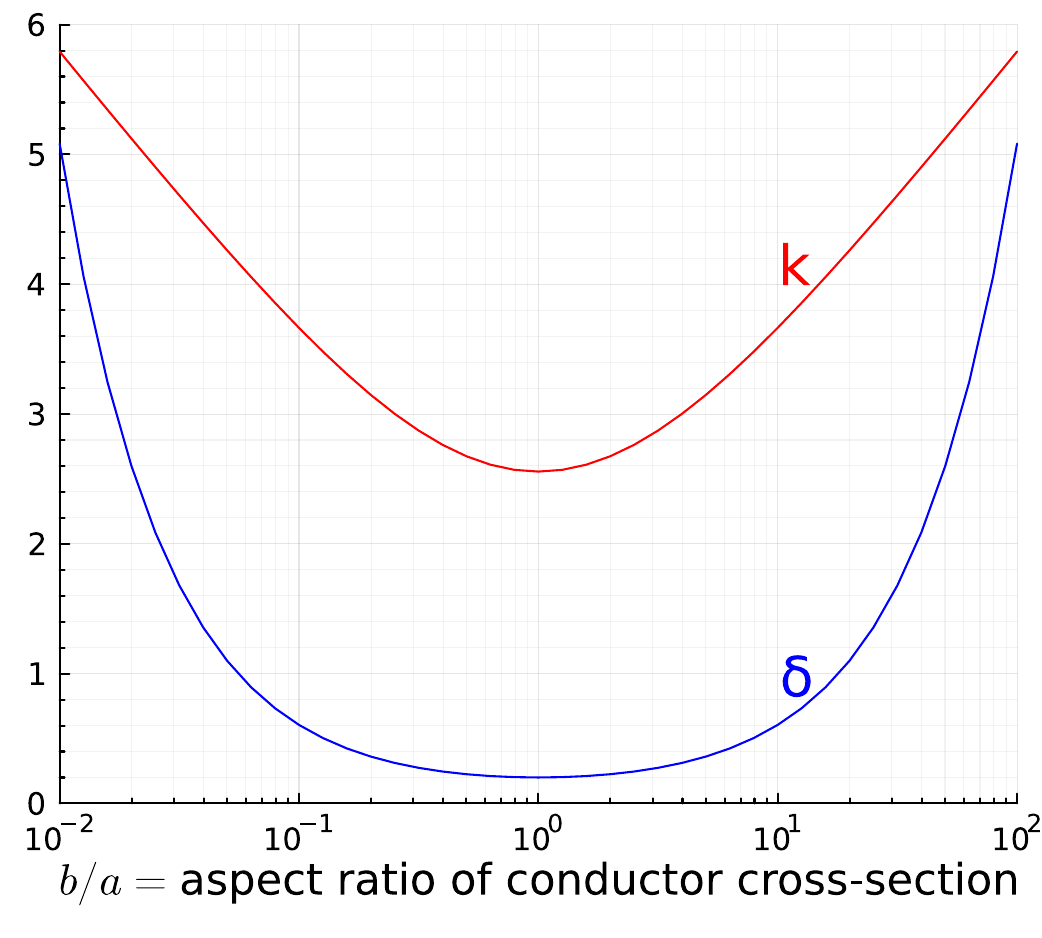}
\caption{The dimensionless quantities $k$ and $\delta$ defined in (\ref{eq:delta})-(\ref{eq:k}) appearing in the reduced models.}
\label{fig:k_delta}
\end{figure}

For convenience, our main results are collected in this section.
These expressions are derived in later sections of the paper.
The self-inductance $L$ is given by
\begin{equation}
\label{eq:L_result}
L =
\frac{\mu_0}{4\pi} \int_0^{2\pi}d\phi \int_0^{2\pi}d\tilde{\phi}
\frac{\mathbf{r}_c' \cdot \tilde{\mathbf{r}}_c'}{\sqrt{ \left| \mathbf{r}_c - \tilde{\mathbf{r}}_c \right|^2 + \delta ab}}, 
\end{equation}
where $\tilde{\mathbf{r}}_c' = d \mathbf{r}_c(\tilde{\phi})/d\tilde{\phi}$,
\begin{equation}
\delta =\exp\left(-\frac{25}{6}+k \right),
\label{eq:delta}
\end{equation}
\begin{IEEEeqnarray}{rCl}
k&=&\frac{4b}{3a}\tan^{-1}\frac{a}{b}+\frac{4a}{3b}\tan^{-1}\frac{b}{a}
+\frac{b^{2}}{6a^{2}}\ln\frac{b}{a}+\frac{a^{2}}{6b^{2}}\ln\frac{a}{b}
\nonumber
\\ 
&&-\frac{a^{4}-6a^{2}b^{2}+b^{4}}{6a^{2}b^{2}}\ln\left(\frac{a}{b}+\frac{b}{a}\right)
.
\label{eq:k}
\end{IEEEeqnarray}
These last two quantities are plotted as a function of the aspect ratio $b/a$ in figure \ref{fig:k_delta}.
For a cross-section that is square, $k=2.5565$ and $\delta=0.19985$.
For $a \gg b$, $k=7/6+\ln(a/b)$ and $\delta=a/(b e^3)$.
As usual, the stored energy follows immediately from $W = (1/2)LI^2$.
For simplicity, the number of turns $N$ is taken to be one; in the multi-turn case, one merely needs to multiply the inductance expressions throughout this paper by $N^2$.

The self-force per unit length is
\begin{equation}
        \frac{d\mathbf{F}}{d\ell} = I \mathbf{t}\times\mathbf{B}_{reg},
    \label{eq:force_result}
\end{equation}
where
\begin{equation}
    \mathbf{B}_{reg}(\phi) 
    =
    \frac{\mu_0 I}{4\pi} \int_0^{2\pi} d\tilde{\phi}
    \frac{\tilde{\mathbf{r}}_c'\times\left(\mathbf{r}_c-\tilde{\mathbf{r}}_c\right)}
    {\left( \left| \mathbf{r}_c-\tilde{\mathbf{r}}_c\right|^2 +\delta ab \right)^{3/2}}.\label{eq:B_reg}    
\end{equation}
The magnetic field inside the conductor is 
\begin{equation}
    \mathbf{B} = \mathbf{B}_{reg} + \mathbf{B}_{0} + \mathbf{B}_\kappa
    +\mathbf{B}_b,
     \label{eq:B_result}
\end{equation}
where
\begin{IEEEeqnarray}{l}
\label{eq:B0}
    \mathbf{B}_{0} = \frac{\mu_0 I}{4\pi a b} \sum_{s_u,s_v} s_u s_v \left[G\left( b(v-s_v),\,a(u-s_u)\right)\mathbf{q} \right.
    \\
    \hspace{1.3in}\left.-G\left(a(u-s_u),\,b(v-s_v)\right)\mathbf{p} \right],
    \nonumber
    \\
    \label{eq:G}
    G(x,y) = y \tan^{-1}\frac{x}{y} + \frac{x}{2}\ln\left(1+\frac{y^2}{x^2}\right),
    \\
    \mathbf{B}_\kappa = \frac{\mu_0 I}{64\pi} \sum_{s_u,s_v} s_u s_v \mathbf{K}\left(u - s_u,\, v - s_v\right),
    \label{eq:B_kappa}
    \\
    \label{eq:K_def}
    \mathbf{K}\left(U,\, V\right) 
    = 
    - 2UV\left(\kappa_1\mathbf{q}-\kappa_2\mathbf{p}\right) \ln\left(\frac{aU^2}{b} + \frac{bV^2}{a}\right)
    \\
    \nonumber \\
    \hspace{0.2in} + \left(\kappa_2 \mathbf{q}-\kappa_1\mathbf{p}\right)\left( \frac{aU^2}{b} + \frac{bV^2}{a}\right)
    \ln\left(\frac{aU^2}{b} + \frac{bV^2}{a}\right)
    \nonumber \\
    \hspace{0.2in} 
    +\frac{4aU^2 \kappa_2\mathbf{p}}{b} \tan^{-1} \frac{bV}{aU} 
    -\frac{4bV^2 \kappa_1 \mathbf{q}}{a} \tan^{-1} \frac{aU}{bV},
    \nonumber
    \\
    \mathbf{B}_b = \frac{\mu_0 I \kappa\mathbf{b}}{8\pi}\left(4+2\ln 2+\ln\delta\right).
    \label{eq:B_b}
\end{IEEEeqnarray}
In (\ref{eq:B0}), (\ref{eq:B_kappa}), and throughout this paper, the sums are over $s_u=\pm 1$ and $s_v = \pm 1$. 
The term $\mathbf{B}_0$ is the field of an infinite straight wire with the same rectangular cross-section as our original general coil.

In the formula (\ref{eq:B_result}) for $\mathbf{B}$, observe that $\mathbf{B}_0 \sim \mu_0 I / a$, whereas $\mathbf{B}_\kappa$ and $\mathbf{B}_b$ are much smaller: $\sim \mu_0 I \kappa$.
Thus, for most applications, $\mathbf{B}_\kappa$ and $\mathbf{B}_b$ can be neglected.
We nonetheless retain them here because for computing the cross-section-averaged Lorentz force, $\mathbf{B}_0$ gives a vanishing contribution to leading order, so these higher order terms are required.

As anticipated in the introduction, the reduced integrals (\ref{eq:L_result}) and (\ref{eq:B_reg}) have a regularizing term in the denominator, $\delta a b$.
Without this term, which prevents division by zero, the integrals would reduce to the expected integrals for infinitesmally thin coils.
This new term is negligible compared to the adjacent term $|\mathbf{r}_c - \tilde{\mathbf{r}}_c|^2$ in most of the integration region.
However, where the separation between the source and evaluation points $\tilde{\mathbf{r}}_c$ and $\mathbf{r}_c$ is comparable to the dimensions of the cross-section, the regularizing term $\delta ab$ becomes important.

Note that the inductance (\ref{eq:L_result}) and force (\ref{eq:force_result}) are independent of the angle by which the cross-section is oriented.

Finally, we will show that equivalent formulas for computing (\ref{eq:L_result}) and (\ref{eq:B_reg})
which can be more accurate numerically
are
\begin{IEEEeqnarray}{rCl}
 \label{eq:L_result_quadrature}
L & = & \frac{\mu_{0}}{4\pi}\int_0^{2\pi} d\phi\left|\mathbf{r}_{c}'\right|\ln\left(\frac{64}{\delta ab}\left|\mathbf{r}_{c}'\right|^{2}\right)\\
 &  & +\frac{\mu_{0}}{4\pi}\int_0^{2\pi} d\phi\int_0^{2\pi} d\tilde{\phi}\left(\frac{\mathbf{r}_{c}'\cdot\tilde{\mathbf{r}}_{c}'}{\sqrt{\left|\mathbf{r}_{c}-\tilde{\mathbf{r}}_{c}\right|^{2}+\delta ab}}
 \right.
 \nonumber \\
 &&\hspace{0.9in}\left.
 -\frac{\left|\mathbf{r}_{c}'\right|^{2}}{\sqrt{\left[2-2\cos\left(\tilde{\phi}-\phi\right)\right]\left|\mathbf{r}_{c}'\right|^{2}+\delta ab}}\right)
 \nonumber
\end{IEEEeqnarray}
and
\begin{IEEEeqnarray}{rCl}
\label{eq:B_reg_quadrature}
\mathbf{B}_{reg} 
&=& \frac{\mu_{0}I \kappa \mathbf{b}}{8\pi}
\left[-2+\ln\left(\frac{64}{\delta ab}\left|\mathbf{r}_{c}'\right|^{2}\right)\right]
 \\
&& +\frac{\mu_{0}I}{4\pi}\int_{0}^{2\pi}d\tilde{\phi}\left[\frac{\tilde{\mathbf{r}}'_{c}\times\left(\mathbf{r}_{c}-\mathbf{\tilde{r}}_{c}\right)}{\left(\left|\mathbf{r}_{c}-\mathbf{\tilde{r}}_{c}\right|^{2}+\delta ab\right)^{3/2}}
 \right.
 \nonumber \\
&&\left. 
 -\mathbf{r}_{c}'\times\mathbf{r}_{c}'' \frac{1-\cos\left(\tilde{\phi}-\phi\right)}{\left(\left[2-2\cos\left(\tilde{\phi}-\phi\right)\right]\left|\mathbf{r}_{c}'\right|^{2}+\delta ab\right)^{3/2}}\right].
 \nonumber
\end{IEEEeqnarray}
In both (\ref{eq:L_result_quadrature}) and (\ref{eq:B_reg_quadrature}), the last term cancels the near-singularity in (\ref{eq:L_result}) and (\ref{eq:B_reg}), so the integrands are significantly smoother.
Therefore the integrals can be evaluated to reasonable accuracy with many fewer quadrature points.

We now proceed to show how the formulas of this section are derived.


\section{Methodology}
\label{sec:methodology}

To rigorously justify the reduced formulas (\ref{eq:L_result})-(\ref{eq:B_b}), we begin by establishing orderings.
Let $R$ indicate the large spatial scale of the coil center-line, with $R \sim |\mathbf{r}_c'| \sim \kappa^{-1} \sim \kappa_1^{-1} \sim \kappa_2^{-1} \sim \kappa_3^{-1} $.
Ordering $a \sim b$, we consider an expansion in the small parameter $a / R \ll 1$.
Moreover, an intermediate scale $d$ is introduced, satisfying
\begin{equation}
a \ll d \ll R.
\end{equation}
Associated with $d$, we define an angle $\phi_0=d/R$.
In the original high-dimensional integrals for the inductance and field, we separate the integrals into a sum of two parts, depending on whether the source and evaluation $\phi$ locations are within $\phi_0$ or not.
This condition reflects whether the source and evaluation points are within a distance $d$ or not.
In the ``far'' term, where the difference in $\phi$ exceeds $\phi_0$, the finite cross-section of the conductor can be neglected, resulting in a substantial simplification of the integrals.
In the ``near'' term, where the source and evaluation $\phi$ values are within $\phi_0$, the shape of the coil can be Taylor-expanded.
The resulting simplified integrals can then be performed analytically.
Now that the total quantity (inductance or field) has been expressed as a sum of two simplified terms, a reduced 1- or 2-dimensional integral is proposed as an ansatz.
The proposed integral is similarly written as a sum of near and far contributions.
The near parts of the high-fidelity and reduced models are shown to be equal.
Similarly, the far parts of the high-fidelity and reduced models are equal.
Thus, the high-fidelity and reduced models are equivalent in the limit $a / R \ll 1$.

A related method of introducing an intermediate scale has been used previously for inductance problems.
See e.g. page 123 of \cite{landau2013electrodynamics}, or \cite{Dengler}.

We now proceed to apply this procedure to the inductance, force, and field in the following sections.
The force can be computed either directly, by integrating the force density $\mathbf{J}\times\mathbf{B}$, or indirectly by applying the principle of virtual work to the inductance.
The latter turns out to be convenient because it requires only lowest-order terms, whereas the direct calculation of the force requires finding higher-order terms in the field.


\section{Self-inductance and stored magnetic energy}
\label{sec:inductance}


The self-inductance $L$ is defined by equating $(1/2)LI^2$ to the magnetic energy $W=\int d^3r B^2/(2\mu_0)=\int d^3r \mathbf{A}\cdot\mathbf{J}/2$, where $\mathbf{A}$ is the vector potential satisfying $\mathbf{B}=\nabla\times\mathbf{A}$. Thus,
\begin{equation}
L=\frac{1}{I^2} \int d^3r \mathbf{A}\cdot\mathbf{J} ,
\;\;\mbox{where}\;\;
\mathbf{A} = \frac{\mu_0}{4\pi} \int d^3 \tilde{r} 
\frac{\tilde{\mathbf{J}}}{\left| \mathbf{r} - \tilde{\mathbf{r}}\right|}.
\label{eq:L_and_A_def}
\end{equation}
Tildes indicate quantities evaluated at the source location specified by the integration variables.
As explained in the previous section, the total potential is written as a sum of near and far contributions, partitioned in $\phi$. Using (\ref{eq:J}),
\begin{equation}
\mathbf{A}=\mathbf{A}_{near}+\mathbf{A}_{far},
\end{equation}
\begin{equation}
\mathbf{A}_{near}=\frac{\mu_{0}I}{4\pi ab}\int_{-1}^{1}d\tilde{u}\int_{-1}^{1}d\tilde{v}\int_{\phi-\phi_{0}}^{\phi+\phi_{0}}d\tilde{\phi}
\frac{\sqrt{\tilde{g}}\tilde{\mathbf{t}}}{\left|\mathbf{r}-\tilde{\mathbf{r}}\right|},\label{eq:A_near_def}
\end{equation}
\begin{equation}
\mathbf{A}_{far}=\frac{\mu_{0}I}{4\pi ab}\int_{-1}^{1}d\tilde{u}\int_{-1}^{1}d\tilde{v}\int_{\phi+\phi_{0}}^{2\pi+\phi-\phi_{0}}d\tilde{\phi}
\frac{\sqrt{\tilde{g}}\tilde{\mathbf{t}}}{\left|\mathbf{r}-\tilde{\mathbf{r}}\right|}.\label{eq:A_far_def}
\end{equation}
Based on this decomposition, (\ref{eq:L_and_A_def}) gives
the self-inductance as a sum of the two corresponding terms:
\begin{equation}
L=L_{near}+L_{far},
\end{equation}
\begin{equation}
L_{near}=\frac{1}{abI}\int_{-1}^{1}du\int_{-1}^{1}dv\int_{0}^{2\pi}d\phi \, \sqrt{g}\mathbf{A}_{near}\cdot\mathbf{t},\label{eq:L_near_def}
\end{equation}
\begin{equation}
L_{far}=\frac{1}{abI}\int_{-1}^{1}du\int_{-1}^{1}dv\int_{0}^{2\pi}d\phi \,  \sqrt{g}\mathbf{A}_{far}\cdot\mathbf{t}.\label{eq:L_far_def}
\end{equation}
In these last two lines, note that the $\phi$ integrals without tildes
still cover the full interval $[0,2\pi)$.

For the far contribution, $\left|\mathbf{r}-\tilde{\mathbf{r}}\right|\gg d$, so
all of the nonzero-thickness terms can be dropped:
$\mathbf{r}-\tilde{\mathbf{r}}\approx\mathbf{r}_{c}-\tilde{\mathbf{r}_{c}}$.
Also the Jacobian (\ref{eq:Jacobian_exact}) can be approximated as
\begin{equation}
    \sqrt{g} \approx \frac{ab \left| \mathbf{r}_c' \right| }{4} .
    \label{eq:Jacobian_approx}
\end{equation}
The integrals (\ref{eq:A_far_def}) and (\ref{eq:L_far_def}) then reduce to
\begin{equation}
\mathbf{A}_{far}=\frac{\mu_{0}I}{4\pi}\int_{\phi+\phi_{0}}^{2\pi+\phi-\phi_{0}}\frac{d\tilde{\phi} \; \tilde{\mathbf{r}}_c'}{\left|\mathbf{r}_{c}-\tilde{\mathbf{r}}_{c}\right|},\label{eq:A_far}
\end{equation}
\begin{equation}
L_{far}=\frac{\mu_{0}}{4\pi}\int_{0}^{2\pi}d\phi\int_{\phi+\phi_{0}}^{2\pi+\phi-\phi_{0}}d\tilde{\phi}\frac{\mathbf{r}_c' \cdot \tilde{\mathbf{r}}_c'}{\left|\mathbf{r}_{c}-\tilde{\mathbf{r}}_{c}\right|}.\label{eq:L_far_hifi}
\end{equation}

To evaluate $\mathbf{A}_{near}$ and $L_{near}$, quantities in (\ref{eq:A_near_def}) are Taylor-expanded about $\phi' \approx \phi$.
Details of the evaluation of 
$\mathbf{A}_{near}$ and $L_{near}$ are given in appendix \ref{sec:A_L_near}.
After a lengthy calculation, the result is
\begin{equation}
    L_{near} = \frac{\mu_0}{4\pi}\int_0^{2\pi}d\phi
    \left| \mathbf{r}_c'\right|
     \left[ 
    2 \ln\left(\frac{2 \phi_0 | \mathbf{r}_c'|}{\sqrt{ab}}\right) 
    +\frac{25}{6}- k
    \right],
    \label{eq:L_near_hifi}
\end{equation}
with $k$ given by (\ref{eq:k}).

We now propose an ansatz of the form (\ref{eq:L_result}), where $\delta$ remains to be determined.
Calling this expression $L^{fil}$, we can write it without approximation as a similar sum of two terms:
\begin{equation}
L^{fil}=L_{far}^{fil}+L_{near}^{fil},
\end{equation}
\begin{equation}
L_{near}^{fil}=\frac{\mu_{0}}{4\pi}\int_{0}^{2\pi}d\phi\int_{\phi-\phi_{0}}^{\phi+\phi_{0}}d\tilde{\phi}\frac{\mathbf{r}_{c}'\cdot\tilde{\mathbf{r}}_{c}'}{\sqrt{\left|\mathbf{r}_{c}-\tilde{\mathbf{r}}_{c}\right|^{2}+\delta ab}},
\end{equation}
\begin{equation}
L_{far}^{fil}=\frac{\mu_{0}}{4\pi}\int_{0}^{2\pi}d\phi\int_{\phi+\phi_{0}}^{2\pi+\phi-\phi_{0}}d\tilde{\phi}\frac{\mathbf{r}_{c}' \cdot \tilde{\mathbf{r}}_{c}'}{\sqrt{\left|\mathbf{r}_{c}-\tilde{\mathbf{r}}_{c}\right|^{2}+\delta ab}}.
\end{equation}
In the far term, the $\delta ab$ term in the denominator is negligible compared to $\left|\mathbf{r}_{c}-\tilde{\mathbf{r}}_{c}\right|^{2}$.
Then comparing with (\ref{eq:L_far_hifi}), we see that $L^{fil}_{far}=L_{far}$.
In $L^{fil}_{near}$, approximating $\left|\mathbf{r}_{c}-\tilde{\mathbf{r}}_{c}\right|^{2} \approx (\tilde{\phi}-\phi)^2 |\mathbf{r}'_c|^2$, the integral can be evaluated using (\ref{eq:phi_integral_approx}), with the result
\begin{equation}
    L^{fil}_{near} = \frac{\mu_0}{4\pi}\int_0^{2\pi}d\phi\, \left|\mathbf{r}_c'\right|
    \left[ 
    2 \ln\left(\frac{2 \phi_0 \left|\mathbf{r}_c'\right|}{\sqrt{ab}}\right) 
    - \ln\delta
    \right].
\end{equation}
Comparing this expression to (\ref{eq:L_near_hifi}), we see that $L^{fil}_{near}=L_{near}$ if we choose $\delta$ according to (\ref{eq:delta}).
Thus, the high-fidelity calculation of $L$ by (\ref{eq:L_and_A_def}) rigorously agrees with (\ref{eq:L_result}) in the limit $a \ll R$.

\begin{figure}[!t]
\centering
\includegraphics[width=\columnwidth]{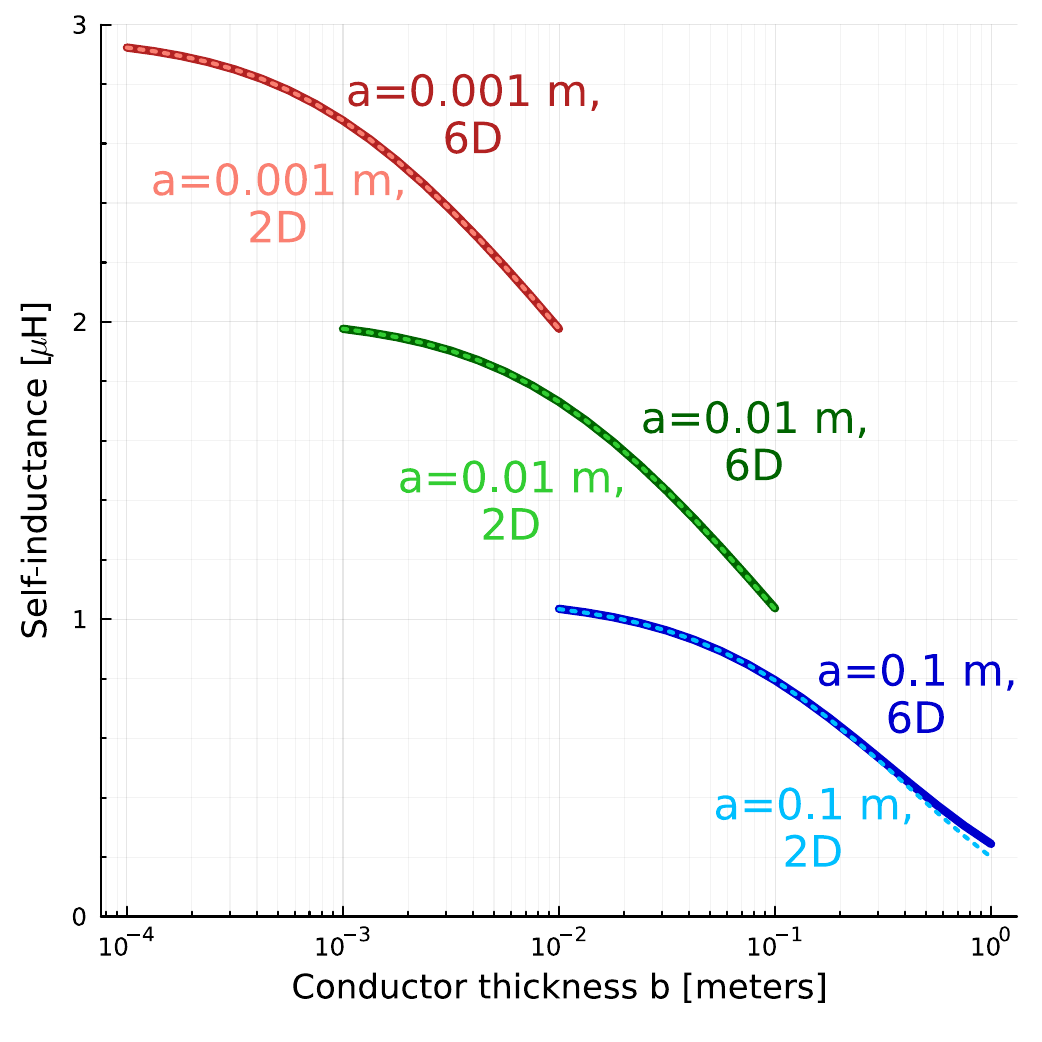}
\caption{The self-inductance of the HSX coil in figure \ref{fig:3D}, computed with the 2D reduced model (\ref{eq:L_result}) (dashed curves), closely agrees with high-fidelity calculations using the 6D integral (\ref{eq:inductance_6D}) (solid curves).
The agreement is excellent across a wide range of values for the conductor cross-section dimensions $a$ and $b$.
}
\label{fig:inductance}
\end{figure}

We can now confirm the accuracy of the reduced model (\ref{eq:L_result}) by numerical comparison to the high-fidelity 6D integral (\ref{eq:inductance_6D}).
Throughout the paper, such comparisons between high-fidelity and reduced models will be made using the HSX coil displayed in figure \ref{fig:3D}.
The rectangular cross-section is chosen to be oriented along the ``coil centroid frame'' described in \cite{Singh}.
This frame is defined in terms of the centroid $\mathbf{C} = (1/l) \int d\ell \;\mathbf{r}_c$, where $d\ell = |\mathbf{r}_c'| d\phi$ is the incremental arclength and $l=\int d\ell$ is the total length.
The frame is defined by
\begin{equation}
    \mathbf{p} = \frac{\mathbf{w} - (\mathbf{w}\cdot\mathbf{t})\mathbf{t}}{\left|\mathbf{w} - (\mathbf{w}\cdot\mathbf{t})\mathbf{t}\right|},
    \hspace{0.3in}
    \mathbf{q}=\mathbf{t}\times\mathbf{p},
\end{equation}
where $\mathbf{w}=\mathbf{r}_c - \mathbf{C}$.
Thus, $\mathbf{p}$ is a vector pointing outward from the centroid, projected to be orthogonal to the tangent.
Here, and in later sections of the paper, the multi-dimensional integrals are evaluated using the \texttt{HCubature.jl} package, a Julia implementation of the algorithm in \cite{genz1980remarks}.
Code used for all the numerical computations in this paper can be obtained at
\cite{github} and
\cite{zenodo}.

Figure \ref{fig:inductance} shows a comparison of the reduced model (\ref{eq:L_result}) with the high-fidelity 6D integral (\ref{eq:inductance_6D}) for the HSX coil.
To test the dependence on $a$ and $b$ in (\ref{eq:delta})-(\ref{eq:k}), the aspect ratio of the cross-section $b/a$ is varied from 0.1 to 10 for three values of $a$.
For comparison, the length of the coil is $l=2.05$ m, comparable to a circle with radius $l/(2\pi)=0.327$ m.
The figure shows that the reduced model agrees extremely well with the high-fidelity model.
Differences are imperceptible on the scale of the plot except at the largest values of $a$ and $b$, where the expansion is not as accurate.

If the coil consists of $N$ turns, with current $I_N = I/N$ per turn, the current density and hence magnetic energy $W$ is unchanged.
The inductance $L_N$ in this case is defined by $W=(1/2) L_N I_N^2$, so $L_N = N^2 L$, hence the inductance formulas throughout the paper are multiplied by $N^2$.

In the case of a circular coil, the inductance expressions here for both the high-fidelity and reduced models simplify to a classical formula.
This result is shown in appendix \ref{sec:circle}.


\section{Internal field}
\label{sec:field}

Now we derive the model for the magnetic field inside the conductor, (\ref{eq:B_result})-(\ref{eq:B_b}).
As explained in section \ref{sec:methodology}, we will proceed by showing that the near and far parts of the high fidelity model match those of the reduced model.
There are two options for computing $\mathbf{B}$: either compute $\mathbf{A}$ to the requisite order and then take the curl, or directly evaluate the Biot-Savart law (\ref{eq:BiotSavart_3D}).
Here we will follow the latter approach.
(To find $\mathbf{B}$ to the desired accuracy from $\mathbf{A}$, $\mathbf{A}$ would need to be computed to higher order than the calculation in Appendix \ref{sec:A_L_near}.)

Similar to section (\ref{sec:inductance}), we begin by writing the high-fidelity expression (\ref{eq:BiotSavart_3D}) as a sum of near and far contributions, partitioned in $\phi$:
\begin{equation}
\label{eq:B_hifi_sum}
\mathbf{B}=\mathbf{B}_{near}+\mathbf{B}_{far},
\end{equation}
\begin{equation}
\label{eq:B_near_def}
    \mathbf{B}_{near}
    =\frac{\mu_{0}I}{4 \pi ab}\int_{-1}^{1}d\tilde{u}\int_{-1}^{1}d\tilde{v}\int_{\phi-\phi_{0}}^{\phi+\phi_{0}}d\tilde{\phi}
     \frac{\sqrt{\tilde{g}}\tilde{\mathbf{t}}\times (\mathbf{r}-\tilde{\mathbf{r}})}{\left|\mathbf{r}-\tilde{\mathbf{r}}\right|^3},
\end{equation}
\begin{equation}
\label{eq:B_far_def} 
\mathbf{B}_{far}
=\frac{\mu_{0}I}{4 \pi ab}\int_{-1}^{1}d\tilde{u}\int_{-1}^{1}d\tilde{v}\int_{\phi+\phi_{0}}^{2\pi+\phi-\phi_{0}}\hspace{-0.1in}d\tilde{\phi}
\frac{\sqrt{\tilde{g}}\tilde{\mathbf{t}}\times (\mathbf{r}-\tilde{\mathbf{r}})}{\left|\mathbf{r}-\tilde{\mathbf{r}}\right|^3}.
\end{equation}

In the far term, the source and evaluation points are at least a distance $\sim d \gg a$ apart, so the thickness of the conductor can be neglected.
Hence we can approximate $\mathbf{r} \approx \mathbf{r}_c$ and use (\ref{eq:Jacobian_approx}), with the result
\begin{equation}
\mathbf{B}_{far}
=\frac{\mu_{0}I}{4\pi}
\int_{\phi+\phi_{0}}^{2\pi+\phi-\phi_{0}}d\tilde{\phi}
\left|\tilde{\mathbf{r}}_{c}'\right|
\frac{\tilde{\mathbf{t}}\times (\mathbf{r}_c-\tilde{\mathbf{r}}_c)}{\left|\mathbf{r}_c-\tilde{\mathbf{r}}_c\right|^3}.
 \label{eq:B_far}
\end{equation}

\begin{figure*}[!t]
\centering
\raisebox{.45\height}{\includegraphics[height=1.4in]{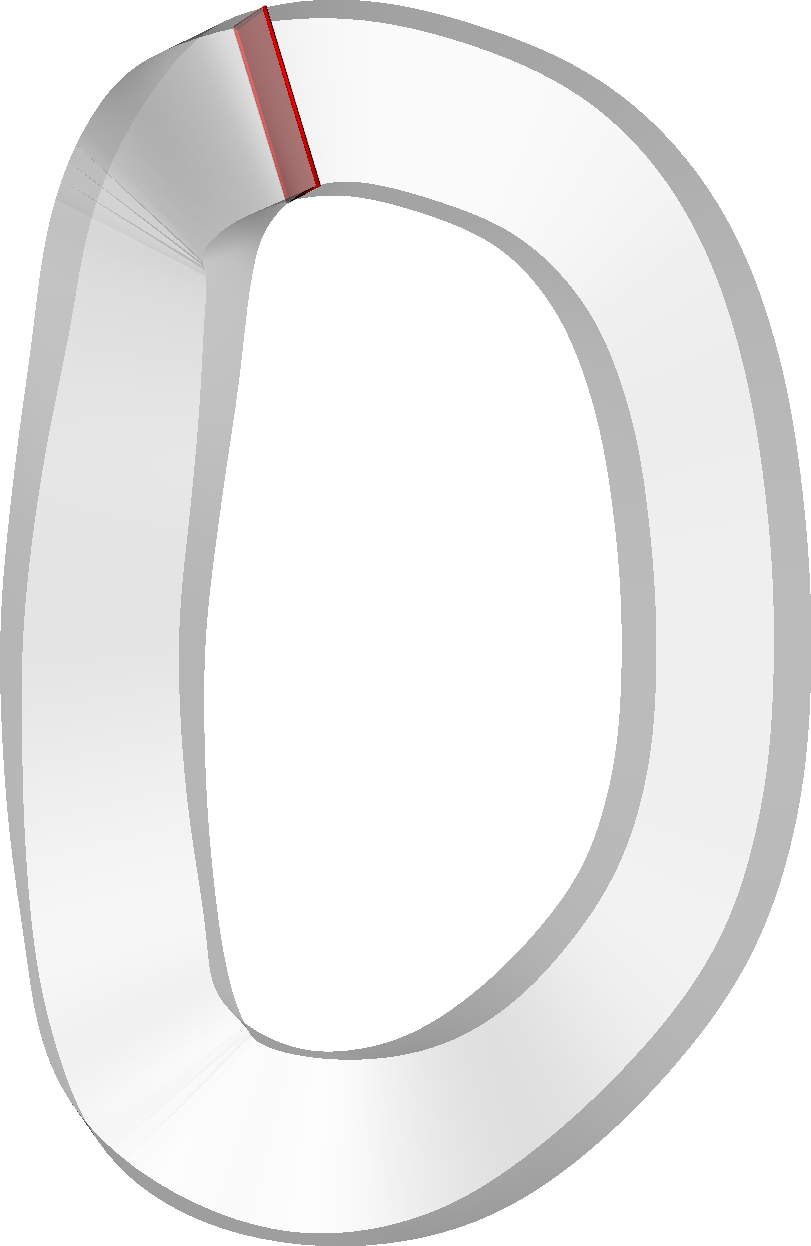}}
\includegraphics[height=2.4in]{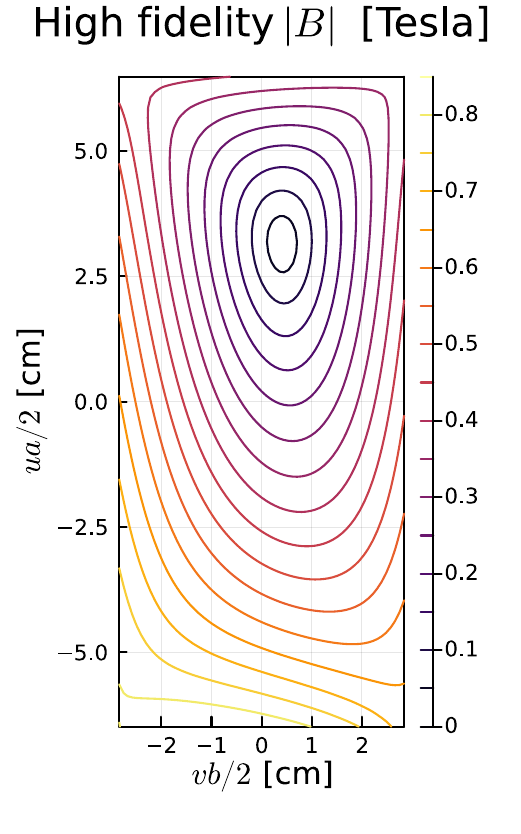}
\includegraphics[height=2.4in]{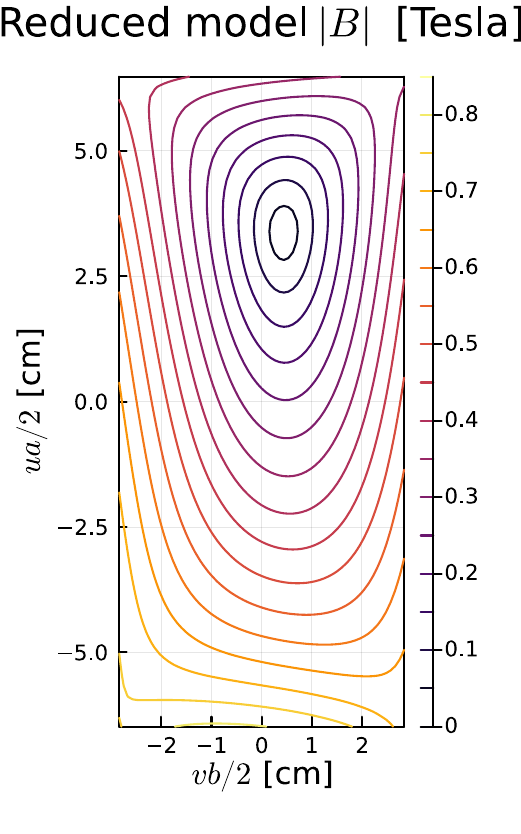}
\includegraphics[height=2.4in]{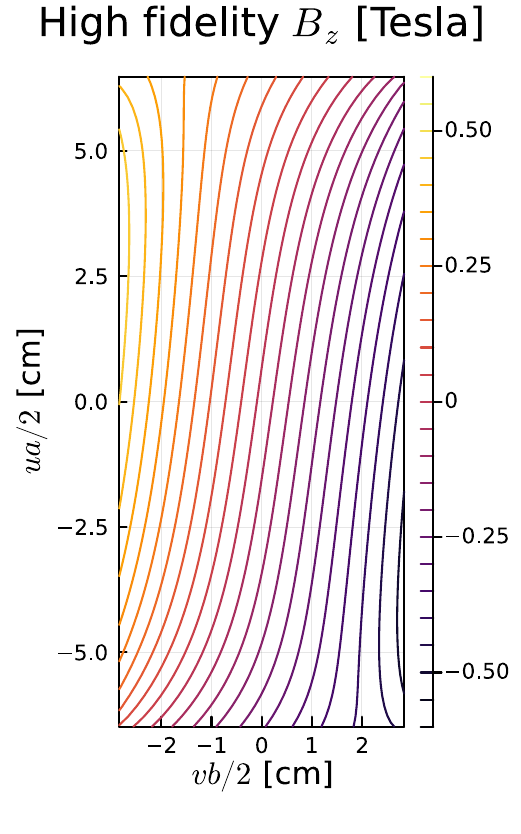}
\includegraphics[height=2.4in]{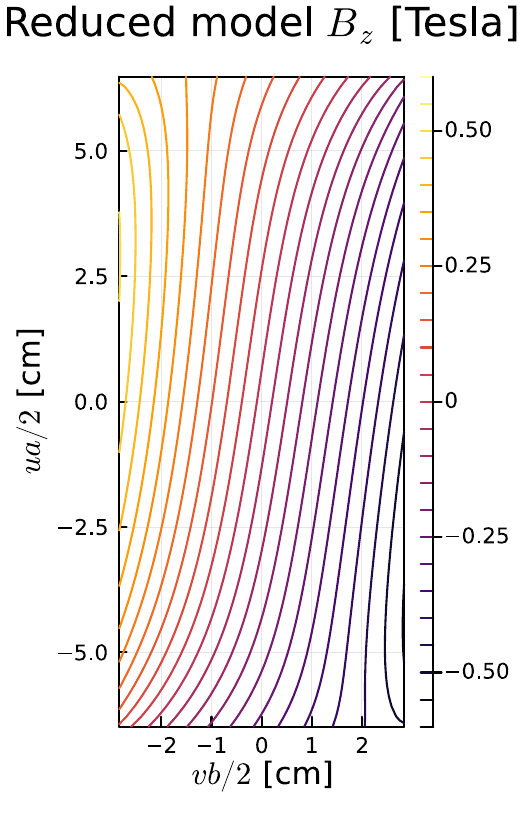}
\caption{
Magnetic field inside the conductor of the HSX coil, for the cross-section highlighted in red at left.
The reduced model (\ref{eq:B_result}), requiring only a 1D integral, agrees well with the high fidelity model - the 3D integral (\ref{eq:BiotSavart_3D}).
Both the field magnitude and one component ($z$) are shown.
}
\label{fig:field}
\end{figure*}

To evaluate $\mathbf{B}_{near}$, quantities in (\ref{eq:B_near_def}) are Taylor-expanded about $\tilde{\phi} \approx \phi$.
Details are relegated to Appendix \ref{sec:field_details}.
The result obtained is
\begin{equation}
\label{eq:B_near_result}
    \mathbf{B}_{near} = \mathbf{B}_0 + \mathbf{B}_{\kappa} 
    + \frac{\mu_0 I \kappa \mathbf{b}}{4\pi} \left[ 1 + \ln\left(\frac{4 \phi_0 |\mathbf{r}'_c|}{\sqrt{ab}}\right)\right] 
\end{equation}
where $\mathbf{B}_0$ is the field of an infinite straight wire in (\ref{eq:B0}), and $\mathbf{B}_{\kappa}$ is given by (\ref{eq:B_kappa}).

Now that the far and near behavior of the high-fidelity model is determined, we can guess a form for a reduced model with only a 1D integral.
In contrast to the inductance and force, the near internal field depends on $u$ and $v$, which would not be captured by a 1D integral alone like (\ref{eq:L_result}) or (\ref{eq:force_result}).
Therefore we must add this local variation.
A natural ansatz is
\begin{equation}
    \mathbf{B} = \mathbf{B}_{reg} + \mathbf{B}_{uv},
    \label{eq:B_ansatz}
\end{equation}
for some $\mathbf{B}_{uv}(\phi,u,v)$ to be determined, where $\mathbf{B}_{reg}$ is given by (\ref{eq:B_reg}).
We can write $\mathbf{B}_{reg}=\mathbf{B}^{reg}_{near}+\mathbf{B}^{reg}_{far}$ where
\begin{IEEEeqnarray}{l}
\label{eq:B_reg_near}
    \mathbf{B}^{reg}_{near}
    =
    \frac{\mu_0 I}{4\pi} \int_{\phi-\phi_0}^{\phi+\phi_0} d\tilde{\phi}
    \frac{\tilde{\mathbf{r}}_c'\times\left(\mathbf{r}_c-\tilde{\mathbf{r}}_c\right)}
    {\left( \left| \mathbf{r}_c-\tilde{\mathbf{r}}_c\right|^2 +\delta ab \right)^{3/2}},
    \\
    \mathbf{B}^{reg}_{far}
    =
    \frac{\mu_0 I}{4\pi} \int_{\phi+\phi_0}^{2\pi+\phi-\phi_0} d\tilde{\phi}
    \frac{\tilde{\mathbf{r}}_c'\times\left(\mathbf{r}_c-\tilde{\mathbf{r}}_c\right)}
    {\left( \left| \mathbf{r}_c-\tilde{\mathbf{r}}_c\right|^2 +\delta ab \right)^{3/2}}.
\label{eq:B_reg_far}
\end{IEEEeqnarray}
In $\mathbf{B}^{reg}_{far}$, the $\delta ab$ in the denominator can be neglected compared to $|\mathbf{r}_c - \tilde{\mathbf{r}}_c|^2$, so $\mathbf{B}^{reg}_{far} \approx \mathbf{B}_{far}$.
To evaluate $\mathbf{B}^{reg}_{near}$, we approximate $\mathbf{r}_c-\tilde{\mathbf{r}}_c \approx (\phi-\tilde{\phi}) \mathbf{r}_c'$ in the denominator, and use
\begin{equation}
    \tilde{\mathbf{r}}_c'\times\left(\mathbf{r}_c-\tilde{\mathbf{r}}_c\right)
    \approx
    \frac{1}{2}(\tilde{\phi} - \phi)^2 |\mathbf{r}_c'|^3 \kappa\mathbf{b}
\end{equation}
in the numerator.
Then the integral may be evaluated using (\ref{eq:phi_integral_32_2}), with the result
\begin{equation}
\label{eq:B_reg_near_result}
    \mathbf{B}^{reg}_{near} = \frac{\mu_0 I \kappa\mathbf{b}}{8\pi}
    \left[-2+2\ln\left(\frac{2\phi_0|\mathbf{r}_c'|}{\sqrt{ab}}\right)-\ln\delta\right].
\end{equation}
Equating (\ref{eq:B_hifi_sum}) with (\ref{eq:B_ansatz}), the filament model and high-fidelity calculation agree if one chooses
\begin{equation}
    \mathbf{B}_{uv}=\mathbf{B}_0 + \mathbf{B}_\kappa + \frac{\mu_0 I \kappa \mathbf{b}}{8\pi} \left( 4 + 2\ln 2+\ln\delta\right).
\end{equation}
Eq (\ref{eq:B_result}) follows.
Getting the reduced model for $\mathbf{B}$ to match the high-fidelity calculation did not require any particular value for $\delta$ so it could be set to 1 for simplicity, but then $\mathbf{B}_{reg}$ could not be re-used for the force calculation (\ref{eq:force_result}), since the particular value of $\delta$ does turn out to matter in that context.

The reduced model derived here is demonstrated in figure \ref{fig:field}.
Here, the field inside the conductor is shown for the HSX coil.
In the the left panel of figure \ref{fig:field}, the cross-section is shown as a red rectangle.
The field magnitude and one field component are displayed in the right four panels, both for the reduced model (\ref{eq:B_result}), and for the original 3D Biot-Savart integral (\ref{eq:BiotSavart_3D}).
The current is taken to be 150 kA.
The high-fidelity and reduced models agree well.
For coils in which the cross-section dimensions are smaller relative to the coil curvature, the agreement improves further.


\section{Self-force}
\label{sec:force}

The self-force can be computed from the self-inductance using the principle of virtual work, as explained in section 3 of \cite{GarrenChen}.
A perturbation $\Delta\mathbf{r}$ to the curve shape is considered, with the current varying to keep the flux $\Phi = LI$ fixed so there is no induced voltage.
The change to the energy $W=(1/2)LI^2$ is then equated to the mechanical work $\int d\ell \,\Delta\mathbf{r}_c \cdot d\mathbf{F}/d\ell$, with the result
\begin{equation}
\int d\ell \, \Delta\mathbf{r}_c \cdot \frac{d\mathbf{F}}{d\ell}
=\frac{I^2}{2} \Delta L.
\label{eq:virtual_work}
\end{equation}
This same result can also be derived by considering the shape perturbation $\Delta\mathbf{r}_c$ to be at fixed current. 
In this case, in addition to the mechanical work, the energy varies also due to the time integral of the $IV$ power associated with the voltage $V = d \Phi / dt$ since the flux varies.

Thus, we consider a perturbation to $\mathbf{r}_c$ in the inductance formula (\ref{eq:L_result}): $\mathbf{r}_c \to \mathbf{r}_c + \Delta \mathbf{r}_c$.
For the terms in which the perturbation acts on the tilde position vector, we can swap the tilde and non-tilde variables.
Integrating by parts to collect factors of $\Delta \mathbf{r}_c$, we obtain
\begin{equation}
\Delta L=\int d\ell\;\Delta\mathbf{r}_c\cdot\left(\mathbf{t}\times\frac{\mu_{0}}{2\pi}\int_{0}^{2\pi}\frac{d\tilde{\phi}\; \tilde{\mathbf{r}}_c' \times\left(\mathbf{r}_c-\tilde{\mathbf{r}}_c\right)}{\left(\left|\mathbf{r}_c-\tilde{\mathbf{r}}_c\right|^{2}+\delta a b\right)^{3/2}}\right).
\label{eq:Delta_L}
\end{equation}

Equation (\ref{eq:virtual_work}) with (\ref{eq:Delta_L}) must hold for any perturbation $\Delta\mathbf{r}_c\left(\ell\right)$, implying the integrands are equal.
Thus, the force formula (\ref{eq:force_result}) with (\ref{eq:B_reg}) follows.

We can also show the same result (\ref{eq:force_result}) can be obtained by integrating the Lorentz force density over the conductor cross-section.
In this approach we proceed as discussed in section \ref{sec:methodology}: the high fidelity calculation is divided into near and far regions, and the reduced model (\ref{eq:force_result}) with (\ref{eq:B_reg}) is shown to give the same near and far contributions.
Beginning with the volume-integrated Lorentz force
$\mathbf{F} = \int d^3r \mathbf{J}\times\mathbf{B}$, applying $d/d\ell = |\mathbf{r}'_c|^{-1} d/d\phi$ gives the 5D integral for the high-fidelity force per unit length:
\begin{IEEEeqnarray}{rCl}
\label{eq:force_5D}
    \frac{d\mathbf{F}}{d\ell} 
    &=& \frac{\mu_0 I^2}{16\pi ab} \int_{-1}^1 du \int_{-1}^1 dv \int_{-1}^1 d\tilde{u} \int_{-1}^1 d\tilde{v} \int_{0}^{2\pi} d\tilde{\phi} 
    \\
    &&\times\left(1 - \frac{ua\kappa_1}{2} - \frac{vb\kappa_2}{2}\right) \sqrt{\tilde{g}}
    \frac{\mathbf{t} \times \left[\tilde{\mathbf{t}} \times (\mathbf{r} - \tilde{\mathbf{r}})\right]}{\left| \mathbf{r} - \tilde{\mathbf{r}}\right|^3}.
    \nonumber
\end{IEEEeqnarray}
Taking advantage of our earlier decomposition (\ref{eq:B_hifi_sum})-(\ref{eq:B_far_def}), we have
\begin{equation}
    \frac{d\mathbf{F}}{d\ell} = \frac{d\mathbf{F}}{d\ell}_{near} + \frac{d\mathbf{F}}{d\ell}_{far}
\end{equation}
where
\begin{equation}
\label{eq:F_near_def}
        \frac{d\mathbf{F}}{d\ell}_{near} 
    =
    \frac{I}{4}\int_{-1}^1 du \int_{-1}^1 dv
    \left(1 - \frac{ua\kappa_1}{2} - \frac{vb\kappa_2}{2}\right)\mathbf{t}\times\mathbf{B}_{near},
\end{equation}
\begin{IEEEeqnarray}{rCl}
    \frac{d\mathbf{F}}{d\ell}_{far} 
    &=&
    \frac{I}{4}\int_{-1}^1 du \int_{-1}^1 dv
    \left(1 - \frac{ua\kappa_1}{2} - \frac{vb\kappa_2}{2}\right)\mathbf{t}\times\mathbf{B}_{far}
    \nonumber
    \\
    &=&
    \frac{\mu_0 I^2}{4\pi}\int_{\phi+\phi_0}^{2\pi+\phi-\phi_0} d\tilde{\phi} |\tilde{\mathbf{r}}'_c|
    \frac{\mathbf{t}\times \left[ \tilde{\mathbf{t}} \times (\mathbf{r}_c - \tilde{\mathbf{r}}_c) \right]}{\left| \mathbf{r}_c - \tilde{\mathbf{r}}_c \right|^3}.
\label{eq:F_far_def}
\end{IEEEeqnarray}
The integral (\ref{eq:F_near_def}) is done in appendix \ref{sec:force_details}, with the result
\begin{equation}
\label{eq:force_near_result}
            \frac{d\mathbf{F}}{d\ell}_{near} 
    =
    \frac{\mu_0 I^2 \kappa \mathbf{n}}{4\pi} \left[ \frac{k}{2} - \frac{13}{12}- \ln\left( \frac{2\phi_0 | \mathbf{r}'_c|}{\sqrt{ab}}\right)\right],
\end{equation}
where $k$ is again given by (\ref{eq:k}).
Next, we consider an ansatz for the force of the form (\ref{eq:force_result}) with (\ref{eq:B_reg}), but not assuming any particular value for $\delta$.
Dividing up the range of the $\phi$ integral as usual, the 1D filament model for the force per unit length is
\begin{equation}
    \frac{d\mathbf{F}}{d\ell}^{fil} = \frac{d\mathbf{F}}{d\ell}_{near}^{fil} + \frac{d\mathbf{F}}{d\ell}_{far}^{fil}
\end{equation}
where
\begin{equation}
    \frac{d\mathbf{F}}{d\ell}_{near}^{fil} = I \mathbf{t}\times \mathbf{B}_{near}^{reg},
    \hspace{0.3in}
    \frac{d\mathbf{F}}{d\ell}_{far}^{fil} = I \mathbf{t}\times \mathbf{B}_{far}^{reg},
\end{equation}
and $\mathbf{B}_{near}^{reg}$ and $\mathbf{B}_{far}^{reg}$ are given in (\ref{eq:B_reg_near}) and (\ref{eq:B_reg_far}).
In $\mathbf{B}_{far}^{reg}$, the $\delta ab$ in the denominator can be neglected compared to $|\mathbf{r}_c - \tilde{\mathbf{r}}_c|^2$, so we find
$(d\mathbf{F}/d\ell)^{fil}_{far} = (d\mathbf{F}/d\ell)_{far}$.
Using (\ref{eq:B_reg_near_result}) in $(d\mathbf{F}/d\ell)^{fil}_{near}$, we find that $(d\mathbf{F}/d\ell)^{fil}_{near}=(d\mathbf{F}/d\ell)_{near}$ if we choose $\delta$ in (\ref{eq:B_reg}) according to (\ref{eq:delta}).
Thus, the near part of the high-fidelity model equals the near part of the reduced model, and the far part of the high-fidelity model equals the far part of the reduced model.
As discussed in appendix \ref{sec:force_details}, this direct approach to justifying the reduced model required high-order terms in $\mathbf{B}_{near}$, whereas the virtual work approach required only the leading-order vector potential via the inductance.

\begin{figure}[!t]
\centering
\includegraphics[width=\columnwidth]{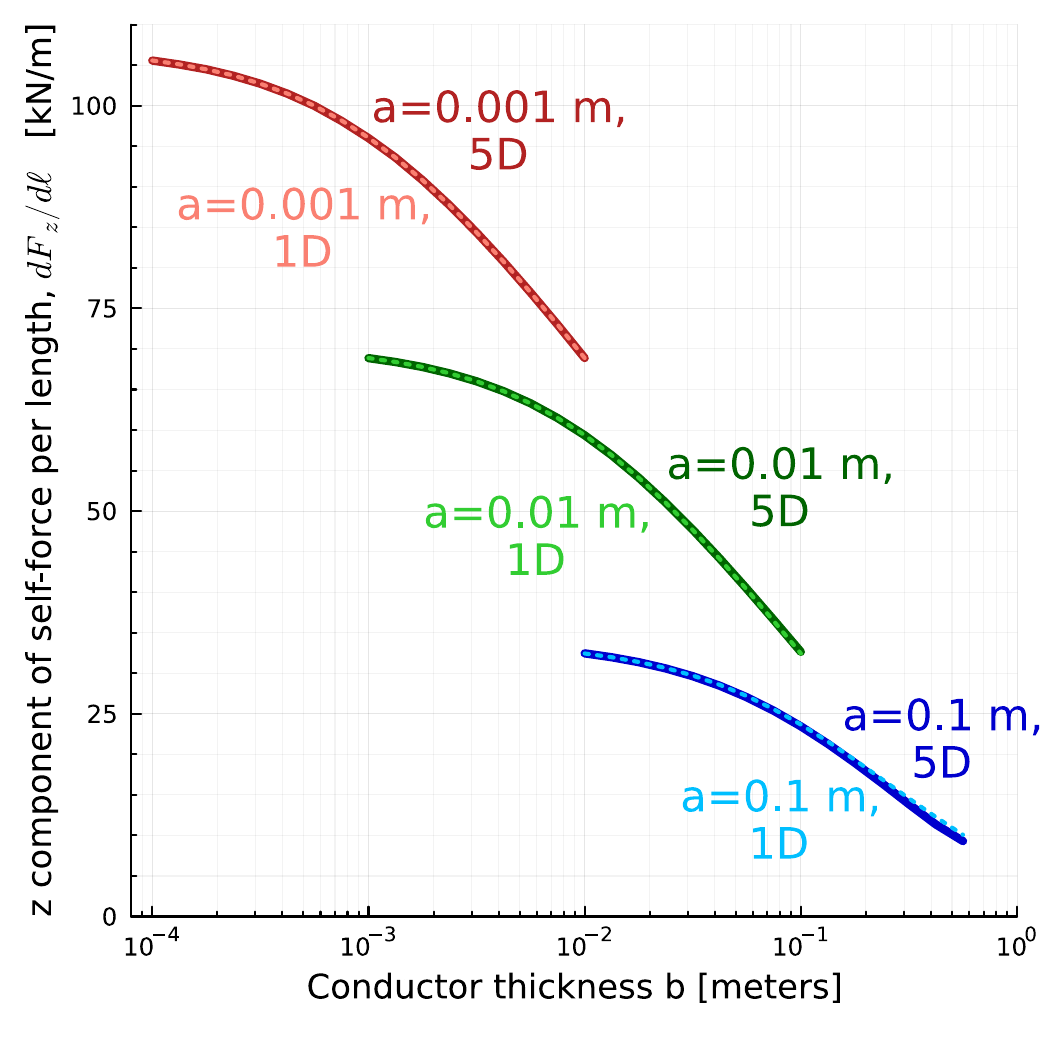}
\caption{The self-force of the HSX coil, computed with the 1D reduced model (\ref{eq:force_result}) (dashed curves), closely agrees with high-fidelity calculations using the 5D integral (\ref{eq:force_5D}) (solid curves).
The $z$ component of the force per unit length is displayed, averaging over the red cross-section shown in figure \ref{fig:field}.
The agreement is excellent across a wide range of values for the conductor cross-section dimensions $a$ and $b$.
}
\label{fig:force_phi0}
\end{figure}

\begin{figure}[!t]
\centering
\includegraphics[width=\columnwidth]{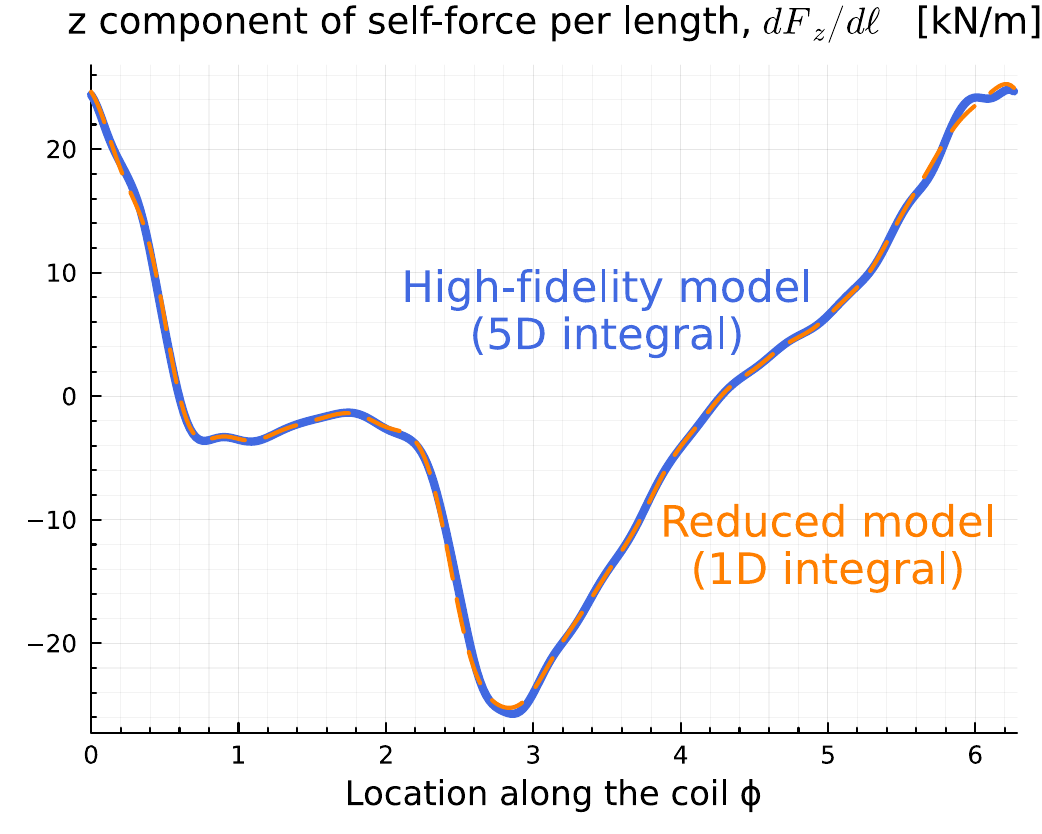}
\caption{Another demonstration that the self-force of the HSX coil, computed with the 1D reduced model (\ref{eq:force_result}) (dashed curve), closely agrees with high-fidelity calculations using the 5D integral (\ref{eq:force_5D}) (solid curve).
Here the true cross-section dimensions of the HSX coil are used.
The $z$ component of the force per unit length is displayed, averaging over the cross-section.
The agreement is excellent at all points along the coil.
}
\label{fig:force_vs_phi}
\end{figure}

The reduced 1D expression for the force agrees quantitatively with the high fidelity 5D integral (\ref{eq:force_5D}) in numerical calculations.
One demonstration of this agreement is shown in figure \ref{fig:force_phi0}.
Here, the two models are evaluated for the HSX coil at the red cross-section highlighted in figure \ref{fig:field}.
Again, the current is taken to be 150 kA.
A particular component ($z$) of the self-force is plotted in figure \ref{fig:force_phi0} for a range of cross-section dimensions $a$ and $b$.
Differences between the high-fidelity and reduced models are imperceptible on the scale of the plot except for the thickest cross-section, for which the $a \ll R$ expansion is less accurate.

A second demonstration of the accuracy of the reduced model is presented in figure \ref{fig:force_vs_phi}.
Here, the cross-section dimensions from figure (\ref{fig:3D}) are used: $a=$ 13 cm and $b=$ 6 cm, matching the real HSX experiment.
In figure \ref{fig:force_vs_phi}, a component of the self-force is plotted at all points along the coil, computed using both the high-fidelity and reduced models.
Even for this thick conductor, the agreement is excellent.


\section{Efficient quadrature}
\label{sec:singularity_subtraction}


\begin{figure}[!t]
\centering
\includegraphics[width=\columnwidth]{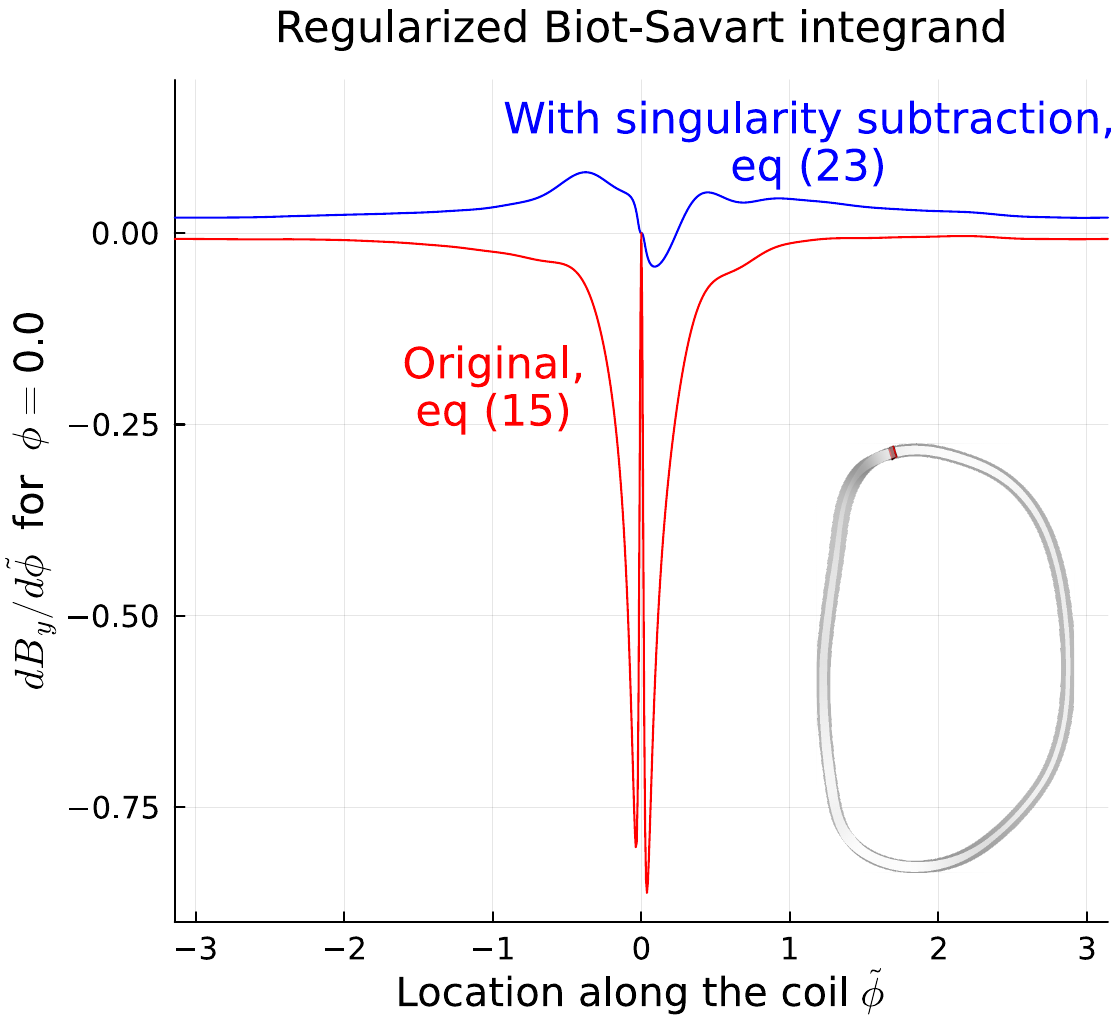}
\caption{Integrand for $\mathbf{B}_{reg}$ without and with the singularity subtraction method of section \ref{sec:singularity_subtraction}.
The coil geometry is shown in the 3D inset, with the red cross-section indicating where $\mathbf{B}_{reg}$ is being evaluated.
Singularity subtraction removes the fine structure from the integrand, easing the resolution required for accurate integration.
}
\label{fig:integrand}
\end{figure}

\begin{figure}[!t]
\centering
\includegraphics[width=\columnwidth]{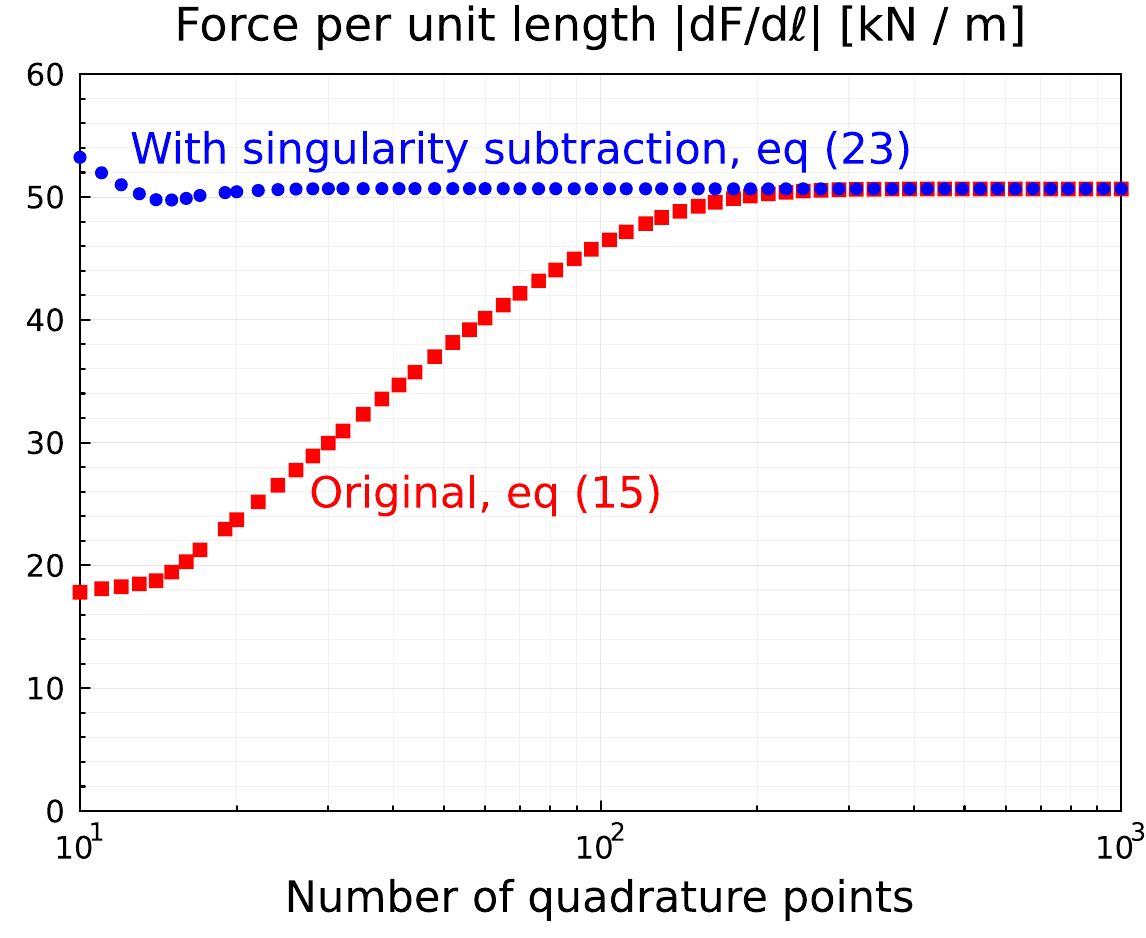}
\caption{Convergence of the self-force (\ref{eq:force_result}) at one point along the coil ($\phi=0$) with respect to the number of quadrature points, without and with the singularity subtraction method of section \ref{sec:singularity_subtraction}.
With singularity subtraction, comparable accuracy can be obtained with many fewer quadrature points.
}
\label{fig:quadrature}
\end{figure}

Here we explain the derivation of  (\ref{eq:L_result_quadrature}) and (\ref{eq:B_reg_quadrature}).
Although in sections \ref{sec:methodology}-{\ref{sec:force}} we have reduced the integrals to two or one dimension, the integrals (\ref{eq:L_result}) and (\ref{eq:B_reg}) still require high resolution in the remaining dimension(s) due to fine structure in the integrands in the nearly singular region.
This structure can be seen in figure \ref{fig:integrand}. 
Here, the integrand for one component of $\mathbf{B}_{reg}$ is plotted for the HSX coil at the point highlighted in figure \ref{fig:field}.
The dimensions $a$ and $b$ are reduced to 2 cm since the issue becomes more pronounced for thin cross-sections.
The fine structure in the integrand is apparent near $\tilde{\phi} = 0$.
Further computational efficiency can therefore be gained by manipulating the reduced integrals (\ref{eq:L_result}) and (\ref{eq:B_reg}) to remove the fine-scale structure.
We proceed by expanding the integrands of (\ref{eq:L_result}) and (\ref{eq:B_reg})  about the nearly singular point.
A term with the same behavior which can be integrated analytically is then subtracted and added.
This calculation is similar to the one in \cite{Hurwitz}.

Beginning with the inductance, and expanding $\tilde{\mathbf{r}}_c$ about $\tilde{\phi}\approx\phi$, the integrand in (\ref{eq:L_result}) is
\begin{equation}
\frac{\mathbf{r}_{c}'\cdot\tilde{\mathbf{r}}_{c}'}{\sqrt{\left|\mathbf{r}_{c}-\tilde{\mathbf{r}}_{c}\right|^{2}+\delta ab}}\approx\frac{\left|\mathbf{r}_{c}'\right|^{2}}{\sqrt{\left(\tilde{\phi}-\phi\right)^{2}\left|\mathbf{r}_{c}'\right|^{2}+\delta ab}}.
\end{equation}
To make the quantity on the right periodic, we replace
\begin{equation}
\left(\tilde{\phi}-\phi\right)^{2}\to 2-2\cos\left(\tilde{\phi}-\phi\right)\label{eq:periodic}
\end{equation}
since the right and left expressions are the same to second order
near $\tilde{\phi}\approx\phi$.
When the result is subtracted and added to (\ref{eq:L_result}), one of the terms is integrated analytically using
\begin{equation}
\int_{0}^{2\pi}\frac{d\phi}{\left(2-2\cos\phi+\Delta\right)^{1/2}}  
=  \frac{4}{\sqrt{4+\Delta}} K\left(\frac{4}{4+\Delta}\right)
\approx\ln\frac{64}{\Delta}
\end{equation}
where $K$ is the complete elliptic integral, and $\Delta = \delta ab / |\mathbf{r}_c'|^2 \ll 1$.
Eq (\ref{eq:L_result_quadrature}) follows.

Next, we proceed in a similar way for $\mathbf{B}_{reg}$.
Expanding $\tilde{\mathbf{r}}_c$ about $\tilde{\phi}\approx\phi$, the integrand in (\ref{eq:B_reg}) is
\begin{equation}
\frac{\tilde{\mathbf{r}}'_{c}\times\left(\mathbf{r}_{c}-\mathbf{\tilde{r}}_{c}\right)}{\left(\left|\mathbf{r}_{c}-\mathbf{\tilde{r}}_{c}\right|^{2}+\delta ab\right)^{3/2}}\approx\frac{\frac{1}{2}\left(\tilde{\phi}-\phi\right)^{2}\mathbf{r}_{c}'\times\mathbf{r}_{c}''}{\left[\left(\tilde{\phi}-\phi\right)^{2}\left|\mathbf{r}_{c}'\right|^{2}+\delta ab\right]^{3/2}}.
\label{eq:B_reg_integrand}
\end{equation}
Again, (\ref{eq:periodic}) is applied to make the quantity on the right of (\ref{eq:B_reg_integrand}) periodic.
When the result is subtracted and added to (\ref{eq:B_reg}), one of the terms is integrated analytically using
\begin{IEEEeqnarray}{l}
\label{eq:circle_phi_integral1_Delta}
\int_{0}^{2\pi}d\phi\frac{2-2\cos\phi}{\left(2-2\cos\phi+\Delta\right)^{3/2}}
\\
=-\frac{4}{\sqrt{4+\Delta}}\left[E\left(\frac{4}{4+\Delta}\right)-K\left(\frac{4}{4+\Delta}\right)\right]
\approx-2+\ln\frac{64}{\Delta}
\nonumber
\end{IEEEeqnarray}
for $\Delta \ll 1$, where $E$ and $K$ are complete elliptic integrals.
Using $\mathbf{r}_c'\times\mathbf{r}_c''/|\mathbf{r}_c'|^3=\kappa\mathbf{b}$, we obtain (\ref{eq:B_reg_quadrature}).

The blue curve in figure \ref{fig:integrand} shows how this singularity-subtraction method removes the fine-scale structure from the integrand for $\mathbf{B}_{reg}$.
We therefore expect the number of quadrature points required for accurate integration to be substantially reduced.
This expectation is borne out in figure \ref{fig:quadrature}, which shows the convergence of the self-force with respect to the number of quadrature points.
The coil parameters are the same as in figure \ref{fig:integrand}, and uniformly spaced quadrature points in $\tilde{\phi}$ with uniform weights are used.
It is clear that the singularity-subtraction method greatly improves the convergence.
Similar behavior is observed for the inductance (not shown).

For a comparison of the computational savings that can be gained by the reduced models in this paper, we have computed the self-force per unit length for the point along the coil in figure \ref{fig:field}, both by integrating over the full finite-thickness cross-section  (eq (\ref{eq:force_5D}) and using our reduced 1D model (eq (\ref{eq:force_result})-(\ref{eq:B_reg})).
Adaptive quadrature is used for the 5D integral, while uniformly spaced quadrature points are used for the 1D integral.
Both formulas are evaluated in Julia on one core of a 2020 M1 Macbook laptop.
For three significant digits in the result, the high fidelity calculation takes 1.0 s, whereas the reduced model takes $57\; \mu$s, a factor of $\sim$ 18,000 speed-up.


\section{Conclusion}
\label{sec:conclusion}

In this work we have derived reduced expressions for efficiently calculating several quantities of interest for electromagnets: the self-inductance, magnetic energy, field inside the conductor, and self-force per unit length.
These quantities normally must be computed by numerical calculations that resolve the cross-section of the conductor, involving high-dimensional integrals or equivalent partial differential equations.
Here, by using analytic methods to reduce the problem, the quantities can be computed using only 1D integrals (for the self-field and self-force) or 2D integrals (for the self-inductance and energy).
The main results have been collected in section \ref{sec:results}.
The reduced models were formulated to rigorously match the original high-fidelity formulas in the limit that the conductor cross-section dimensions were small compared to the curvature of the coil center-line.
Interestingly, the inductance, energy, and force-per-unit-length are found to be independent of the angle by which the rectangular cross-section is oriented.
Throughout the paper, the reduced models have been compared to high-fidelity calculations in which the conductor cross-section is fully resolved.
The agreement is excellent even for complicated coil shapes such as that in figure \ref{fig:3D}.

A key assumption of this work was that the current density was uniform across the cross-section.
This approximation is accurate if there are many turns of the conductor in both dimensions of the cross-section.
However the uniform-current-density approximation might not be sufficiently accurate  in superconducting coils for some quantities of interest.
For example, the distribution of current across a high-temperature-superconducting tape is typically nonuniform \cite{rostila2007self, gomory2006self},
determined by physics not included here.
The critical current of a superconducting coil is a property of practical importance, and precise computation of the critical current likely requires determination of this nonuniform current distribution by a method as in \cite{riva2023development}.
However, the internal field computed by (\ref{eq:B_result}) might at least provide an estimate for the critical current, as follows.
First the coil current $I$ is initialized to a plausible value; $I$ will be updated later.
Given $\mathbf{B}$ from (\ref{eq:B_result}), a formula for the local critical current density $j_c$ in terms of the local field (such as eq (6) of \cite{gomory2006self}) gives the local critical current density.
This current density could be integrated over the cross-section to get a critical current at each $\phi$, and minimization over $\phi$ would then give an estimate for the overall critical current $I_c$:
\begin{equation}
\label{eq:critical_current}
    I_c \approx  \min_{\phi} \frac{ab}{4}
    \int_{-1}^1 du \int_{-1}^1 dv \;j_c(\mathbf{B}(\phi,u,v,I)).
\end{equation}
Root-finding in one variable could then be applied to solve for the coil current $I$ such that $I_c=I$, consistent with (\ref{eq:critical_current}).
This approach corresponds to the first iteration of the self-consistent numerical solution in \cite{rostila2007self,gomory2006self}.

One other limitation of the work here is that sharp corners in the coil center-line are disallowed.
It might be possible to address this case in future research.

In this work, the force per unit length was computed by averaging the Lorentz force over the cross-section.
However other quantities related to Lorentz forces may also be of interest, including local force variation within the conductor cross-section, and stresses associated with gradients of the local force density.
These additional quantities can be computed also using the approach here, since the full magnetic field vector within the conductor has been derived, (\ref{eq:B_result}).

A potential application for the results of this paper is in design optimization of electromagnets.
One might wish to optimize the shapes of magnets for reduced peak forces, reduced magnetic energy, or reduced peak field.
In optimization, it is necessary to evaluate the objective function many times, so speed of evaluation is a priority.
Therefore the efficient formulas here will be advantageous.


%

\appendices
\section{Details of the near vector potential and self-inductance calculation}
\label{sec:A_L_near}


Now we evaluate $\mathbf{A}_{near}$. 
Again (\ref{eq:Jacobian_approx}) can be used.
In (\ref{eq:A_near_def}), $\tilde{\mathbf{r}}_c'$ and $\tilde{\mathbf{t}}$ can be approximated by the corresponding quantities at the non-tilde (evaluation) point.
In the denominator, for most of the integration domain, $\phi-\tilde{\phi} \sim \phi_0 = d/R$, in which case $\left|\mathbf{r}-\tilde{\mathbf{r}}\right| \approx \left|\phi-\tilde{\phi} \right| |\mathbf{r}_c'|$.
However if this expression is used in (\ref{eq:A_near_def}), the integral diverges near $\phi \approx \phi_0$.
Evidently, more care is required in an inner layer where $\phi-\tilde{\phi} \sim a/R$. In this layer, we have to leading order
\begin{equation}
    \left|\mathbf{r}-\tilde{\mathbf{r}}\right|^2 \approx
    \left(\phi-\tilde{\phi} \right)^2 \left|\mathbf{r}_c'\right|^2
    +\frac{a^2}{4}(u-\tilde{u})^2 + \frac{b^2}{4}(v-\tilde{v})^2.
    \label{eq:delta_r}
\end{equation}
This expression is valid also in the rest of the near integration region where $\phi-\tilde{\phi} \sim \phi_0$. Thus,
\begin{IEEEeqnarray}{l}
    \mathbf{A}_{near}=
    \frac{\mathbf{t}\mu_{0}I}{16\pi}\left|\mathbf{r}_{c}'\right|\int_{-1}^{1}d\tilde{u}\int_{-1}^{1}d\tilde{v}\int_{\phi-\phi_{0}}^{\phi+\phi_{0}}d\tilde{\phi} 
    \\
   \times \left[\left(\tilde{\phi}-\phi\right)^{2}\left|\mathbf{r}_{c}'\right|^{2}+\frac{a^{2}}{4}\left(u-\tilde{u}\right)^{2}+\frac{b^{2}}{4}\left(v-\tilde{v}\right)^{2}\right]^{-1/2}.
   \nonumber
\end{IEEEeqnarray}
The $\tilde{\phi}$ integral is done using
\begin{IEEEeqnarray}{l} 
\int_{-\phi_{0}}^{\phi_{0}}d\phi\frac{1}{\left(x+y\phi^{2}\right)^{1/2}}
\nonumber
\\
=\frac{1}{\sqrt{y}}\ln\left(1+\frac{2\phi_{0}\left(y\phi_{0}+\sqrt{y}\sqrt{x+y\phi_{0}^{2}}\right)}{x}\right)
\nonumber
\\
\approx\frac{1}{\sqrt{y}}\ln\left(\frac{4y\phi_{0}^{2}}{x}\right),\label{eq:phi_integral_approx}
\end{IEEEeqnarray}
where $x/(y\phi_0^2) \sim a^2/d^2 \ll 1$ has been used.
The remaining $\tilde{u}$ and $\tilde{v}$ integrals can be done using
\begin{IEEEeqnarray}{l}
\label{eq:uv_integral}
\int_{-1}^1 d\tilde{u} \int_{-1}^1 d\tilde{v} \, \ln\left(\frac{a}{b}\left(u-\tilde{u}\right)^2
+\frac{b}{a}\left(v-\tilde{v}\right)^2\right)
\\
=-12+4\sum_{s_u,s_v} s_u s_v F\left(u-s_u,\, v-s_v\right)
\nonumber
\end{IEEEeqnarray}
where the sum is over $s_{u}, s_v \in\left\{ -1,1\right\} $ and
\begin{IEEEeqnarray}{rCl}
\label{eq:F}
F\left(U,V\right)
&=&\frac{aU^{2}}{4b}\tan^{-1}\frac{bV}{aU}+\frac{bV^{2}}{4a}\tan^{-1}\frac{aU}{bV}
\\
&&+\frac{UV}{4}\ln\left(\frac{aU^{2}}{b}+\frac{bV^{2}}{a}\right).
\nonumber
\end{IEEEeqnarray}
To derive (\ref{eq:uv_integral}, 
the $\tilde{u}$ integral is straightforward, but
the $\tilde{v}$ integral that follows requires care because terms in the integrand are non-smooth at $\tilde{v}=v$.
These terms are correctly integrated using
\begin{IEEEeqnarray}{l}    
\int_{-1}^{1}d\tilde{v}\,\left(v-\tilde{v}\right)\tan^{-1}\frac{c}{v-\tilde{v}}
 \\
=
c
+\sum_{s_v} \frac{s_v}{2} \left[c^2\tan^{-1}\frac{v-s_v}{c}
-\left(v-s_v\right)^{2}\tan^{-1}\frac{c}{v-s_v}\right],
\nonumber
\end{IEEEeqnarray}
for $s_v \in \{-1, 1\}$ and any $c$. Finally we obtain
\begin{IEEEeqnarray}{l}
\label{eq:A_near_result}
\mathbf{A}_{near}=\frac{\mathbf{t}\mu_{0}I}{4\pi}\left[3+\ln\left(\frac{16\phi_{0}^{2} \left| \mathbf{r}_c'\right|^2}{ab}\right)
\right.
\\
\hspace{0.98in}
\left.-\sum_{s_{u}, s_v}s_{u}s_{v}F\left(u-s_{u}, v-s_{v}\right)\right].
\nonumber
\end{IEEEeqnarray}

With $\mathbf{A}_{near}$ now determined,
the $u$ and $v$ integrals in (\ref{eq:L_near_def}) (with (\ref{eq:Jacobian_approx})) can be done using
\begin{equation}
    \int_{-1}^1 du \int_{-1}^1 dv \sum_{s_u,s_v} s_u s_v
    F(u-s_u, v-s_v)
    =
    8\ln 2 - \frac{14}{3} + 4k.
\end{equation}
The result is (\ref{eq:L_near_hifi}).


\section{Details of the near magnetic field calculation}
\label{sec:field_details}


Here we give details of the calculation of $\mathbf{B}_{near}$.
We first evaluate the leading-order contributions to (\ref{eq:B_near_def}).
In the denominator, (\ref{eq:delta_r}) can be used, as in appendix \ref{sec:A_L_near}.
In the numerator of (\ref{eq:B_near_def}), we have to leading order
\begin{equation}
        \tilde{\mathbf{t}}\times(\mathbf{r}-\tilde{\mathbf{r}}) 
        \approx
    (u-\tilde{u})\frac{a}{2}\mathbf{q}-(v-\tilde{v})\frac{b}{2}\mathbf{p}
+\frac{\kappa}{2}(\tilde{\phi}-\phi)^2\left|\mathbf{r}_c'\right|^2 \mathbf{b},
\end{equation}
where (\ref{eq:kappa_b}) has been used.
The terms on the right-hand side have a relative ratio of $d^2/(aR)$ which is not obviously large or small.
Using (\ref{eq:Jacobian_approx}) and $\left| \tilde{\mathbf{r}}_c'\right| \approx \left| \mathbf{r}_c'\right|$ for the Jacobian, we have
\begin{equation}
    \mathbf{B}_{near} \approx \mathbf{B}^{(0)} + \mathbf{B}^{(1)}
\end{equation}
where
\begin{IEEEeqnarray}{rCl}
\label{eq:B_sup_0}
    \mathbf{B}^{(0)} 
    &=&\frac{\mu_0 I \left| \mathbf{r}_c'\right| }{32\pi}\int_{-1}^1 d\tilde{u}
    \int_{-1}^1 d\tilde{v} \int_{\phi-\phi_0}^{\phi+\phi_0}d\tilde{\phi}    \\
    && \times \frac{(u-\tilde{u})a\mathbf{q}-(v-\tilde{v})b\mathbf{p}}{\left[
    \left(\phi-\tilde{\phi} \right)^2 \left|\mathbf{r}_c'\right|^2
    +\frac{a^2}{4}(u-\tilde{u})^2 + \frac{b^2}{4}(v-\tilde{v})^2\right]^{3/2}},
    \nonumber \\
\label{eq:B_sup_1}
        \mathbf{B}^{(1)} 
    &=&\frac{\mu_0 I \left| \mathbf{r}_c'\right|^3 \kappa\mathbf{b}}{32\pi}\int_{-1}^1 d\tilde{u}
    \int_{-1}^1 d\tilde{v} \int_{\phi-\phi_0}^{\phi+\phi_0}d\tilde{\phi}    \\
    && \times \frac{\left(\tilde{\phi}-\phi\right)^2  }
    {\left[
    \left(\phi-\tilde{\phi} \right)^2 \left|\mathbf{r}_c'\right|^2
    +\frac{a^2}{4}(u-\tilde{u})^2 + \frac{b^2}{4}(v-\tilde{v})^2\right]^{3/2}}.
    \nonumber
\end{IEEEeqnarray}
In (\ref{eq:B_sup_0}), the $\tilde{\phi}$ integral is first evaluated using 
\begin{equation}
\int_{-\phi_{0}}^{\phi_{0}}d\phi\frac{1}{\left(x+y\phi^{2}\right)^{3/2}}=\frac{2\phi_{0}}
{x\sqrt{x+y\phi_{0}^{2}}}
\approx
\frac{2}{x\sqrt{y}},
\label{eq:phi_integral_32_1}
\end{equation}
for $x\ll y\phi_{0}^{2}$.
Then doing the $\tilde{u}$ and $\tilde{v}$ integrals, we find $\mathbf{B}^{(0)}=\mathbf{B}_0$ with $\mathbf{B}_0$ given by (\ref{eq:B0})-(\ref{eq:G}).
In (\ref{eq:B_sup_1}), the $\tilde{\phi}$ integral is first evaluated using
\begin{IEEEeqnarray}{l}
    \int_{-\phi_{0}}^{\phi_{0}}d\phi\frac{\phi^{2}}{\left(x+y\phi^{2}\right)^{3/2}}
    \nonumber
    \\
    =\frac{2}{y^{3/2}}\left[\frac{\ln x}{2}-\phi_{0}\sqrt{\frac{y}{x+y\phi_{0}^{2}}} \right.
    \nonumber \\
    \hspace{1in}\left.-\ln\left(-\sqrt{y}\phi_{0}+\sqrt{x+y\phi_{0}^{2}}\right)\right]
    \nonumber
    \\
\approx
\frac{1}{y^{3/2}}\left[-2+2\ln2-\ln\left(\frac{x}{y}\right)+2\ln\phi_{0}\right],
\label{eq:phi_integral_32_2}
\end{IEEEeqnarray}
where the last line holds for $x\ll y\phi_{0}^{2}$.
The $\tilde{u}$ and $\tilde{v}$ integrals are performed using (\ref{eq:uv_integral}), and the result is
\begin{IEEEeqnarray}{l}
\label{eq:B_sup_1_result}
    \mathbf{B}^{(1)} = \frac{\mu_0 I \kappa \mathbf{b}}{8\pi}
    \\
 \times   \left[ 1 +
    2\ln\left(\frac{4\phi_0 |\mathbf{r}_c'|}{\sqrt{ab}}\right)
    -\sum_{s_u, s_v} s_u s_v F(u-s_u, \, v-s_v)
    \right],
    \nonumber
\end{IEEEeqnarray}
where the sum is over $s_u,s_v \in \{-1, 1\}$,
and $F$ is given by (\ref{eq:F}).

Observe that $\mathbf{B}^{(0)}$ is $\sim \mu_0 I / a$, whereas $\mathbf{B}^{(1)}$ is formally smaller, with terms of size $\sim \mu_0 I / R$ and $\sim (\mu_0 I/R) \ln(R\phi_0/a)$.
Also $\mathbf{B}_{far}$ has terms of size $\sim \mu_0 I / R$ and $\sim (\mu_0 I/R) \ln\phi_0$, which can be seen by evaluating $\mathbf{B}_{far}$ for a circular coil in (\ref{eq:B_far_circle}).
Let us therefore compute all of the terms in $\mathbf{B}$ through order $\sim \mu_0 I / R$.
Indeed, we will need all terms of this order to compute the self-force by direct integration of $\mathbf{J}\times\mathbf{B}$.
The terms in $\mathbf{B}^{(1)}$ are already small so no high-order corrections are needed there, but corrections do need to be kept in the larger term $\mathbf{B}^{(0)}$.
Notice that any corrections to the integrand with an odd power of $(\phi-\tilde{\phi})$ vanish.
Any corrections to the integrand that are small by a factor $(\phi-\tilde{\phi})^2$ will give a contribution to $\mathbf{B}$ that is smaller by $(a^2/R^2)\ln(\phi_0 R/a)$, which can be seen by comparing (\ref{eq:phi_integral_32_1}) with (\ref{eq:phi_integral_32_2}). 
Hence, we only need to keep corrections to the integrand that are smaller than the leading terms by $~a/R$.
There are no such corrections to the numerator of the Biot-Savart law, $\tilde{\mathbf{t}}\times(\mathbf{r}-\tilde{\mathbf{r}})$, but we do need to keep the corrections to the Jacobian (\ref{eq:Jacobian_exact}), and use the following more accurate expression in the denominator:
\begin{IEEEeqnarray}{rCl}
        \left|\mathbf{r}-\tilde{\mathbf{r}}\right|^2 
        &\approx&
    \left(1 - \frac{\kappa_1 a}{2} (u+\tilde{u}) - \frac{\kappa_2 b}{2} (v+\tilde{v})\right)
    \left(\phi-\tilde{\phi} \right)^2 \left|\mathbf{r}_c'\right|^2
    \nonumber
    \\
    &&
    +\frac{a^2}{4}(u-\tilde{u})^2 + \frac{b^2}{4}(v-\tilde{v})^2.
    \label{eq:delta_r_high_order}
\end{IEEEeqnarray}
Then applying (\ref{eq:phi_integral_32_1}),
\begin{equation}
        \int_{\phi-\phi_0}^{\phi+\phi_0} \frac{d\tilde{\phi} \, \sqrt{\tilde{g}}}{\left| \mathbf{r}-\tilde{\mathbf{r}}\right|^{3}}
    \approx 2ab \frac{
    1+\frac{a \kappa_1}{4}(u-\tilde{u}) + \frac{b \kappa_2}{4} (v-\tilde{v})
    }{
    a^2(u-\tilde{u})^2 + b^2 (v-\tilde{v})^2
    }
\end{equation}
We thus find the complete magnetic field through order $\mu_0 I/R$ is
\begin{equation}
\label{eq:B_near_contributions}
\mathbf{B}_{near} \approx \mathbf{B}^{(0)} + \mathbf{B}^{(1)} + \mathbf{B}^{(0)}_1  
\end{equation}
where the new term is
\begin{IEEEeqnarray}{l}
\mathbf{B}_{1}^{(0)}=\frac{\mu_{0}I}{16\pi}\int_{-1}^{1}d\tilde{u}\int_{-1}^{1}d\tilde{v}\frac{\left(u-\tilde{u}\right)a\mathbf{q}-\left(v-\tilde{v}\right)b\mathbf{p}}{a^{2}\left(u-\tilde{u}\right)^{2}+b^{2}\left(v-\tilde{v}\right)^{2}}
\nonumber
\\
\hspace{1in}\times \left[\left(u-\tilde{u}\right)a\kappa_{1}+\left(v-\tilde{v}\right)b\kappa_{2}\right].
\end{IEEEeqnarray}
Evaluating the $\tilde{u}$ and $\tilde{v}$ integrals, and forming the sum (\ref{eq:B_near_contributions}) with (\ref{eq:B_sup_1_result}), we arrive at (\ref{eq:B_near_result}).


\section{Details of the near self-force calculation}
\label{sec:force_details}

In this appendix some details of the derivation of (\ref{eq:force_near_result}) are presented.
The starting point is (\ref{eq:F_near_def}) with (\ref{eq:B_near_result}).
Note that in $\mathbf{B}_{near}$ in (\ref{eq:B_near_result}), the largest term is $\mathbf{B}_0 \sim \mu_0 I / a$, whereas the other terms are $\sim \mu_0 I / R$.
If the quantity in parentheses in (\ref{eq:F_near_def}) is approximated by 1, $
\mathbf{B}_0$ gives a contribution that integrates to zero by symmetry.
Therefore, the small $ua\kappa_1$ and $vb\kappa_2$ terms in (\ref{eq:F_near_def}) must be kept for $\mathbf{B}_0$.
However the quantity in parentheses in (\ref{eq:F_near_def}) can be approximated by 1 for the other terms in $\mathbf{B}_{near}$.
Thus, we have
\begin{IEEEeqnarray}{l}
\label{eq:F_near_intermediate}
\frac{d\mathbf{F}}{d\ell}_{near}
=
-\frac{\mu_0 I^2 \kappa \mathbf{n}}{4\pi} \left[ 1+\ln\left( \frac{4\phi_0 |\mathbf{r}'_c|}{\sqrt{ab}}\right) \right] 
\\
+\frac{I}{8} \int_{-1}^1 du \int_{-1}^1 dv
\left[-(ua\kappa_1+vb\kappa_2)\mathbf{t}\times\mathbf{B}_0
+2\mathbf{t}\times\mathbf{B}_{\kappa}\right].
\nonumber
\end{IEEEeqnarray}
In the $\mathbf{B}_{\kappa}$ term, the contribution from the second line of (\ref{eq:K_def}) vanishes due to symmetry.
In the $\mathbf{p}$ component, the $vb\kappa_2$ term vanishes due to symmetry, as does the $ua\kappa_1$ term in the $\mathbf{q}$ component.
In the remaining integral, it is sufficient to evaluate just one of the $\mathbf{p}$ or $\mathbf{q}$ components, since the other follows by replacing $u \leftrightarrow v$, $a\leftrightarrow b$, $s_u \leftrightarrow s_v$, and $\kappa_1 \leftrightarrow \kappa_2$. 
The necessary integral is
\begin{IEEEeqnarray}{l}
    \int_{-1}^1 du \int_{-1}^1 dv \sum_{s_u,s_v} s_u s_v \left[
    \frac{a (u-s_u)u}{b} \tan^{-1} \frac{b(v-s_v)}{a(u-s_u)}
    \right. \nonumber \\
    +\frac{b(v-s_v)^2}{2a} \tan^{-1} \frac{a(u-s_u)}{b(v-s_v)}
    \nonumber \\
    + \left.\frac{(3u-s_u)(v-s_v)}{4} \ln\left(
    \frac{a(u-s_u)^2}{b} + \frac{b(v-s_v)^2}{a}\right) \right]
    \nonumber \\
    =-\frac{2}{3}+8\ln 2 + 4k.
\end{IEEEeqnarray}
Applying this result in (\ref{eq:F_near_intermediate}), we arrive at (\ref{eq:force_near_result}).


\section{Circular coils}
\label{sec:circle}


In this section, we show how the formulas in this paper simplify when the coil center-line is a circle.
In this case it will be shown that the sum of near and far contributions to each physical quantity becomes independent of the choice of $\phi_0$, as it should.
It will also be shown that the self-inductance formula reduces to a classical expression.

For a circular coil that is concentric with the $z$ axis, typically the cross-section would be oriented such that $\mathbf{p}=\mathbf{n}=-\mathbf{e}_R$ and $\mathbf{q}=\mathbf{b}=\mathbf{e}_z$, where $\mathbf{e}_R$ and $\mathbf{e}_z$ are cylindrical unit vectors.
Then $a$ and $b$ are the radial and axial extent of the conductor, $\kappa_2=0$, and $\kappa_1=\kappa=1/R_0$.

To evaluate the far contributions to the quantities of interest, 
a necessary integral is
\begin{eqnarray}
&&\int_{\phi_{0}}^{2\pi-\phi_{0}}\frac{\cos\chi \,d\chi}{\sqrt{1-\cos\chi} }  \nonumber \\
&& =  -\frac{4}{\sqrt{1-\cos\phi_{0}}}\left[\ln\left(\tan\frac{\phi_{0}}{4}\right)\sin\frac{\phi_{0}}{2}+\sin\phi_{0}\right]\nonumber \\
 && \approx  2\sqrt{2}\left(-2+2\ln2-\ln\phi_{0}\right),\label{eq:circle_integral}
\end{eqnarray}
where the last line is the leading behavior for $\phi_{0}\ll 1$.

Using (\ref{eq:circle_integral}), the far contribution to the vector potential (\ref{eq:A_far}) can be evaluated as
\begin{equation}
\mathbf{A}_{far}=\frac{\mathbf{t}\mu_{0}I}{2\pi}\left(-2+2\ln2-\ln\phi_{0}\right).
\end{equation}
Adding (\ref{eq:A_near_result}), the total vector potential at leading
order is
\begin{IEEEeqnarray}{l}
    \mathbf{A}_{near}=\frac{\mathbf{t}\mu_{0}I}{4\pi}\left[-1+\ln\left(\frac{256 R_{0}^{2}}{ab}\right)
    \right.\\
\hspace{1in}   \left. -\sum_{s_{u},s_v\in\left\{ -1,1\right\} } s_{u}s_{v}F\left(u-s_{u},v-s_{v}\right)\right].
\nonumber
\end{IEEEeqnarray}
As desired, the $\phi_{0}$ terms have cancelled.

The far contribution to the inductance (\ref{eq:L_far_hifi}) evaluates to
\begin{equation}
L_{far}=\mu_{0}R_{0}\left(-2+2\ln2-\ln\phi_{0}\right),\label{eq:L_far_circle}
\end{equation}
where (\ref{eq:circle_integral}) has been used.
Adding the near contribution (\ref{eq:L_near_hifi}), the total inductance becomes
\begin{equation}
L=\mu_{0}R_{0}\left[\ln\left(\frac{8R_{0}}{\sqrt{ab}}\right)+\frac{1}{12}
-\frac{k}{2}
\right],\label{eq:inductance_circle}
\end{equation}
with $k$ given by (\ref{eq:k}).
This classical result can be found as the leading terms on page 353 of \cite{Weinstein1884}, page 423 of \cite{Lyle1914}, or page 137 of \cite{RosaGrover}.
To show the equivalence,
\begin{equation}
    \tan^{-1}\left(\frac{1}{x}\right)=-\tan^{-1} \left(x\right)+\frac{\pi}{2}\mbox{sign}\left(x\right)
\end{equation}
can be used.

The same result can also be obtained from the reduced model (\ref{eq:L_result}).
A necessary integral is
\begin{IEEEeqnarray}{l}
\int_{0}^{2\pi}d\chi\frac{\cos\chi}{\sqrt{1-\cos\chi+\epsilon}} \\
=\frac{4}{\sqrt{2+\epsilon}}\left[\left(1+\epsilon\right)K\left(\frac{2}{2+\epsilon}\right)-\left(2+\epsilon\right)E\left(\frac{2}{2+\epsilon}\right)\right]
\nonumber \\
\approx\sqrt{2}\left(-4+5\ln2-\ln\epsilon\right),
\nonumber
\end{IEEEeqnarray}
where $K$ and $E$ are the complete elliptic integrals, and the last line gives the leading terms for $\epsilon\ll 1$.
Application of this integral to (\ref{eq:L_result}) in the case of a circle again gives (\ref{eq:inductance_circle}).

For the magnetic field in a circular coil, evaluating the high-fidelity 3D integral, (\ref{eq:B_far}) can be evaluated using
\begin{equation}
    \int_{\phi_0}^{2\pi-\phi_0} \hspace{-4mm}\frac{d\chi}{\sqrt{1-\cos\chi}}
    =2\sqrt{2} \ln\left(\cot\frac{\phi_0}{4}\right)
    \approx 2\sqrt{2} (2 \ln 2 -\ln \phi_0)
\end{equation}
as
\begin{equation}
\label{eq:B_far_circle}
    \mathbf{B}_{far} = \frac{\mu_0 I \mathbf{b}}{4\pi R_0}
    (2\ln 2 - \ln\phi_0).
\end{equation}
Adding (\ref{eq:B_near_result}), the total field is
\begin{equation}
\label{eq:B_circle_result}
    \mathbf{B} = \mathbf{B}_0 + \mathbf{B}_{\kappa} 
    + \frac{\mu_0 I  \mathbf{b}}{4\pi R_0} \left[ 1 + \ln\left(\frac{16 R_0}{\sqrt{ab}}\right)\right] ,
\end{equation}
which is manifestly independent of $\phi_0$.
The same result can be obtained from the filament model (\ref{eq:B_result}).
This can be done by using (\ref{eq:B_reg}) with (\ref{eq:circle_phi_integral1_Delta}) to obtain
\begin{equation}
\label{eq:B_reg_circle}
    \mathbf{B}_{reg} = \frac{\mu_0 I \mathbf{e}_z}{4\pi R_0}
    \left[ \ln\left(\frac{8 R_0}{\sqrt{ab}}\right) + \frac{13}{12} - \frac{k}{2}\right].
\end{equation}
Plugging this expression into (\ref{eq:B_result}), we again find (\ref{eq:B_circle_result})

Finally, the force per unit length for a circular coil can be evaluated for the high-fidelity model by substituting (\ref{eq:B_far_circle}) into (\ref{eq:F_far_def}).
Adding (\ref{eq:force_near_result}), we obtain
\begin{equation}
    \frac{d\mathbf{F}}{d\ell} = \frac{\mu_0 I^2 \mathbf{e}_R}{4\pi R_0}
    \left[ \ln\left(\frac{8 R_0}{\sqrt{ab}}\right) + \frac{13}{12} - \frac{k}{2}\right].
\end{equation}
The same result can be obtained from the reduced model  (\ref{eq:force_result}) by applying (\ref{eq:B_reg_circle}).


\section*{Acknowledgment}

We thank 
Robert Granetz,
Stuart Hudson,
Thomas Kruger, 
Jorrit Lion,
Nicolo Riva,
Felix Warmer,
Yuhu Zhai, and
Caoxiang Zhu
for useful discussions.
We are also grateful to Sarah Weiss and Wrick Sengupta for locating and translating \cite{Weinstein1884}.
This work was supported by a grant from the Simons Foundation (No. 560651, T. A.).
This work was also supported by the U.S. Department of Energy under Contract DE-FG02-93ER54197.

\ifCLASSOPTIONcaptionsoff
  \newpage
\fi



\bibliographystyle{IEEEtran}
\bibliography{coil_force_rectangular_xsection}
\end{document}